\documentclass{amsart}
\usepackage{amsfonts,amssymb}

\newtheorem{theorem}{Theorem}
\theoremstyle{plain}

\newtheorem{corollary}{Corollary}

\newtheorem{definition}{Definition}
\newtheorem{example}{Example}

\newtheorem{lemma}{Lemma}

\newtheorem{proposition}{Proposition}
\newtheorem{remark}{Remark}

\numberwithin{equation}{section}
\input{tcilatex}

\font\fiverm=cmr5
\newcommand{\Ref}[1]{\ref{#1}\rlap{\smash{\hbox{\lower3.7pt\hbox{\fiverm
#1}}}}}
\newcommand{\Label}[1]{\label{#1}\rlap{\smash{\hbox{\lower3.7pt\hbox{\fiverm
#1}}}}}
\newcommand{\beq}{\begin{equation}}
\newcommand{\eeq}{\end{equation}}

\newcommand{\ba}{\begin{array}}
\newcommand{\ea}{\end{array}}
\newcommand{\bt}{\begin{theorem}}
\newcommand{\et}{\end{theorem}}
\newcommand{\bp}{\begin{proposition}}
\newcommand{\ep}{\end{proposition}}
\newcommand{\bl}{\begin{lemma}}
\newcommand{\el}{\end{lemma}}
\newcommand{\bc}{\begin{corollary}}
\newcommand{\ec}{\end{corollary}}
\newcommand{\bi}{\begin{itemize}}
\newcommand{\ei}{\end{itemize}}
\newcommand{\ben}{\begin{enumerate}}
\newcommand{\een}{\end{enumerate}}
\newcommand{\bdf}{\begin{definition}\rm}
\newcommand{\edf}{\end{definition}}
\newcommand{\br}{\begin{remark}\rm}
\newcommand{\er}{\end{remark}}
\newcommand{\bex}{\begin{example}\rm}
\newcommand{\eex}{\end{example}}

\def\pri{\hbox to 10pt{\hfil\hbox to 0.4pt{\vrule height5pt width0.4pt
                 depth0pt}\vrule width5pt height0.4pt depth0pt\hfil}}

\begin{document}
\title[Ground states in complex bodies]{\textbf{Ground states in complex
bodies}}
\author{Paolo Maria Mariano}
\address{DICeA, University of Florence, via Santa Marta 3, I-50139 Firenze
(Italy)}
\email{paolo.mariano@unifi.it}
\author{Giuseppe Modica}
\address{Dipartimento di Matematica Applicata "G. Sansone", University of
Florence, via Santa Marta 3, I-50139 Firenze (Italy)}
\email{giuseppe.modica@unifi.it}
\date{September 28th, 2007}

\begin{abstract}
A unified framework for analyzing the existence of ground states in wide
classes of elastic complex bodies is presented here. The approach makes use
of classical semicontinuity results, Sobolev mappinngs and Cartesian
currents. Weak diffeomorphisms are used to represent macroscopic
deformations. Sobolev maps and Cartesian currents describe the inner
substructure of the material elements. Balance equations for irregular
minimizers are derived. A contribution to the debate about the role of the
balance of configurational actions follows. After describing a list of
possible applications of the general results collected here, a concrete
discussion of the existence of ground states in thermodynamically stable
quasicrystals is presented at the end.
\end{abstract}

\maketitle
\tableofcontents

\section{Introduction}

A prominent influence of the material texture (substructure at different low
scales) on the macroscopic mechanical behavior of bodies is often registered
in common experiments of condensed matter physics. Such an influence is
exerted through inner actions conjugated with substructural changes and
bodies displaying it are called \emph{complex}.

In the standard format of continuum mechanics, each material element is
assigned to a place in space and no direct geometrical information about its
inner substructure is given a priori (see the treatise \cite{Sil}). When
active material complexity occurs, active in the sense that peculiar actions
arise, such a scheme does not permits a direct representation of these
actions: they are power-conjugated with morphological changes inside the
material elements. For this reason one finds reasonable to consider each
material element as a system and to describe its inner geometry at least at
coarse grained level. Descriptors of the substructural morphology are
selected here as elements of a finite-dimensional differentiable manifold,
the manifold of substructural shapes, by following this way the general
model-building framework of the mechanics of complex bodies in \cite{C89}, 
\cite{M02}, \cite{MaSt05} (see also \cite{C85}, \cite{dFM}, \cite{M07}, \cite%
{Se}, \cite{BdLS}, \cite{Ho}).

Here the basic aim is to present a general framework for analyzing the
existence of ground states in wide classes of complex bodies, the ones
covered by the multifield modeling sketched above.

The attention is focused on bodies admitting energies of the type 
\begin{equation}
\mathcal{E}\left( u,\nu \right) :=\int_{\mathcal{B}_{0}}e\left( x,u\left(
x\right) ,F\left( x\right) ,\nu \left( x\right) ,N\left( x\right) \right) 
\text{ }dx  \label{1}
\end{equation}%
where $u$ represents the gross deformation, $F$ its spatial derivative, $\nu 
$ the morphological descriptor of the inner substructure, $N$ its spatial
derivative, $\mathcal{B}_{0}$ the reference gross shape of the body. The
approach is characterized by the use of classical semicontinuity results,
Sobolev mappings and Cartesian currents. In particular, the macroscopic
transplacement field (the gross deformation) is considered as a weak
diffeomorphism while the morphological descriptor map as a Sobolev map or a
Cartesian current. It is shown that in the case in which there are energetic
interactions involving minors of both $F$ and $N$, then $u$ and $\nu $
should be considered as a unique map in the setting of Cartesian currents.

The results extend to complex bodies theorems for simple elastic bodies in 
\cite{B1}, \cite{MTY}, \cite{CN}, \cite{GMS3} and \cite{GMS5} (see also the
critical remarks in \cite{B2}).

The structure of the paper is sketched below.

In Section 2, the representation of gross deformation and substructural
morphology in complex bodies is briefly discussed. The variational principle
governing ground states of non-linear elastic complex bodies is stated in
Section 3. The general path leading to existence results is presented in
Section 4. The possible occurrence of a Lavrentiev gap phenomenon and the
consequences of considering the morphological descriptor maps as Cartesian
currents are also discussed. In Section 5, balance equations associated with
irregular minimizers are derived: the natural representation of actions
associated with horizontal variations is used in absence of tangential
derivatives of transplacement fields and morphological descriptor maps. A
list of possible applications of the general framework is presented in
Section 6. Section 7 includes details about the special case of
quasicrystals: the physics suggests the appropriate functional environment
in which ground states may be found.

\section{Morphology and deformation of complex bodies}

In its primary, abstract sense, a \emph{body} is a collection $\mathfrak{B}$
of material elements, each one considered as a patch of matter made of
entangled molecules or the characteristic piece of some atomic lattice. In
other words it is the smallest piece of matter characterizing the nature of
the material constituting a body. The first problem one tackles in thinking
of bodies is the representation of their morphology, that is the
representation of the set $\mathfrak{B}$. In standard continuum mechanics
such a representation is minimalist: Each material element is considered as
a structureless box, a \emph{monad} in Leibnitz's words,\emph{\ }described
only by the place in space of its centre of mass so that one has a bijective
map $\varphi :\mathfrak{B}\rightarrow \mathcal{E}^{3}$ from $\mathfrak{B}$\
into the three-dimensional Euclidean space $\mathcal{E}^{3}$ and calls $%
\varphi \left( \mathfrak{B}\right) $ a \emph{placement} of the body. $%
\varphi \left( \mathfrak{B}\right) $ is indicated by $\mathcal{B}$ and is
assumed to be a bounded domain with boundary $\partial \mathcal{B}$ of
finite two-dimensional measure, a boundary where the outward unit normal is
defined to within a finite number of corners and edges. In this way one is
`collapsing' the material element at a point and neglects any information
about its internal structure. Of course $\mathcal{E}^{3}$\ can be identified
with $\mathbb{R}^{3}$ once an origin is chosen. A reference place $\mathcal{B%
}_{0}:=\varphi _{0}\left( \mathfrak{B}\right) $, generic points of which are
labeled by $x$, is accepted by $\mathbb{R}^{3}$. For technical purposes it
is convenient to distinguish between the space containing $\mathcal{B}_{0}$
and the one in which all other placements of the body are. To this aim an
isomorphic copy of $\mathbb{R}^{3}$, indicated by $\mathbb{\hat{R}}^{3}$, is
selected: it contains each new place $\mathcal{B}:=\varphi \left( \mathfrak{B%
}\right) $. The generic $\mathcal{B}$ is achieved from $\mathcal{B}_{0}$ by
means of a \emph{transplacement }field (the standard deformation) defined by 
$u:=\varphi \circ \varphi _{0}^{-1}$, with 
\begin{equation*}
\mathcal{B}_{0}\ni x\mapsto u\left( x\right) \in \mathcal{B}\text{.}
\end{equation*}%
A basic assumption is that $u$ is \emph{one-to-one} and \emph{orientation
preserving}, the last requirement meaning that at each $x$ the spatial
derivative $Du$ (the standard \emph{gradient of deformation}) has positive
determinant: $\mathop{\rm det}\nolimits Du\left( x\right) >0$. Commonly the
notation $F:=Du\left( x\right) \in Hom\left( T_{x}\mathcal{B}_{0},T_{u\left(
x\right) }\mathcal{B}\right) \simeq \mathbb{R}^{3}\otimes \mathbb{\hat{R}}%
^{3}=M_{3\times 3}$ is adopted, with $M_{3\times 3}$\ the linear space of $%
3\times 3$ matrices.

The standard picture of the morphology of a continuum body does not contain
any direct information about the morphology of the material texture.
However, in complex bodies, the prototype element is a rather complicated
ensemble of entangled molecules, or, more generally, substructures. The
direct description of the substructural morphology is necessary when
alterations of the substructures generate peculiar inner actions within the
body, actions influencing prominently the gross behavior. The representation
of these actions follows from the representation of the substructural
morphology. To account for the inner shape of the material elements, one may
consider a map $\mathfrak{\varkappa :B}\rightarrow \mathcal{M}$ assigning 
\emph{morphological descriptors} selected within a \emph{finite-dimensional} 
\emph{differentiable manifold} $\mathcal{M}$, called \emph{manifold of
substructural shapes}. Elements of $\mathcal{M}$ furnish, in fact, a rough
description of the essential features of the geometry of the substructure.
Unless required by the theorems below, at a first glance $\mathcal{M}$ is
considered here as abstract as possible so that the results are valid for a
wide class of complex bodies, a family possibly restricted only by the
peculiar assumptions required by specific analyses.

Geometrical structures over the manifold of substructural shapes have often
a precise physical meaning so that they have to be attributed to $\mathcal{M}
$ carefully, according to the specific case analyzed. For example, one may
consider $\mathcal{M}$ endowed with boundary to model effects such as
volumetric transitions. Moreover, when the substructure displays its own
peculiar inertia, the related kinetic energy can be represented in its first
approximation by means of a quadratic form. In this way, the (quadratic)
substructural kinetic energy induces a Riemannian structure on $\mathcal{M}$%
, the metric being assigned by the coefficients of the quadratic form. In
general the substructural kinetic energy may naturally induce only a Finsler
structure over $\mathcal{M}$ or some gauge structure (see \cite{M05}, \cite%
{M07}, \cite{CG}).

A \emph{morphological descriptor map }$\nu =\mathfrak{\varkappa }\circ
\varphi _{0}^{-1}$, 
\begin{equation*}
\mathcal{B}_{0}\ni x\longmapsto \nu \left( x\right) \in \mathcal{M},
\end{equation*}
is defined over $\mathcal{B}_{0}$. It is the Lagrangian representation of
the morphological descriptor field and is assumed to be differentiable. Its
spatial derivative $D\nu $ is indicated by $N:=D\nu \left( x\right) \in
Hom\left( T_{x}\mathcal{B}_{0},T_{\nu \left( x\right) }\mathcal{M}\right)
\simeq \mathbb{R}^{3}\otimes T_{\nu }\mathcal{M}\simeq M_{\dim \mathcal{M}
\times 3}.$

The Eulerian (actual) counterpart of the map $\nu $, indicated by $\nu _{a}$
, is given by $\nu _{a}:=\mathfrak{\varkappa }\circ \varphi ^{-1}=\mathfrak{%
\ \varkappa }\circ \varphi _{0}^{-1}\circ u^{-1}$. Its actual derivative $%
D_{a}\nu _{a}$ is an element of $Hom\left( T_{y}\mathcal{B},T_{\nu
_{a}\left( x\right) }\mathcal{M}\right) $ at each $y\in \mathcal{B}$.

The morphological descriptor $\nu $ and its derivative $N$ may or may not
affect the measures of deformation, depending only on the specific
circumstances engaged. Examples clarify the issue. In fact, in the cases in
which $\nu $ represents a microdisplacement, a microscopic independent
rotation of the substructure or it measures a microdeformation, $N$ enters
the measures of gross deformation depending on the special case envisaged
(see examples in \cite{C89}, \cite{MaSt05}). Contrary, when $\nu $\
represents substructural events not related with changes in length or
microdisplacements, the standard measures of deformation are sufficient.

\section{Energy and the variational principle}

Hyper-elastic complex bodies are considered here. A sufficiently smooth
density $e:=e\left( x,u,F,\nu ,N\right) $ describes the states of
equilibrium locally. Each material element is then considered as a system
closed with respect to the exchange of mass and in energetic contact with
the neighboring fellows (the last circumstance being evidenced by the
presence of the gradients $F$ and $N$ in the list of entries of $e$).

Given $u$ and $\nu $, the global energy $\mathcal{E}\left( u,\nu \right) $
of $\mathcal{B}_{0}$ is then simply 
\begin{equation}
\mathcal{E}\left( u,\nu \right) :=\int_{\mathcal{B}_{0}}e\left( x,u\left(
x\right) ,F\left( x\right) ,\nu \left( x\right) ,N\left( x\right) \right) 
\text{ }dx.  \label{Gen}
\end{equation}%
\emph{Ground states }are minimizers of $\mathcal{E}\left( u,\nu \right) $,
i.e. fields $u$ and $\nu $, selected in appropriate functional classes, that
satisfy the variational principle 
\begin{equation*}
\min_{u,\nu }\mathcal{E}\left( u,\nu \right) .
\end{equation*}

In common cases $\mathcal{E}\left( u,\nu \right) $ splits naturally in the
sum 
\begin{equation}
\mathcal{E}\left( u,\nu \right) =\int_{\mathcal{B}_{0}}e^{i}\left( x,F\left(
x\right) \mathbf{,}\nu \left( x\right) \mathbf{,}N\left( x\right) \right) 
\text{ }dx+\int_{\mathcal{B}_{0}}\left( e_{1}^{e}\left( u\left( x\right)
\right) +e_{2}^{e}\left( \nu \left( x\right) \right) \right) \text{ }dx,
\label{Spec}
\end{equation}
where $e^{i}\left( x,F,\nu ,N\right) $ is the \emph{internal `stored' energy}
, $e_{1}^{e}\left( y\right) $ the potential of standard bulk (gravitational)
forces and $e_{2}^{e}\left( \nu \right) $ the potential of direct bulk
actions over the substructure such as electric fields in the case of
polarizable substructures.

The requirement of objectivity, that is the invariance with respect to the
action of $SO\left( 3\right) $ (here on both the ambient space $\mathbb{\hat{%
R}}^{3}$ and $\mathcal{M}$), implies that in the range of large deformations
the energy density $e\left( x,y,F,\nu ,N\right) $ cannot be convex with
respect to $F$, see pertinent comments in \cite{Sil}), once one fixes the
other arguments, while it may be a convex function of $N$ (see \cite{MaSt05}%
).

A prominent special case of (\ref{Gen}) is the one of \emph{partially
decomposed energies}. They are characterized by two sufficiently smooth
functions $e_{E}$ and $e_{M}$ such that 
\begin{equation}
\mathcal{E}\left( u,\nu \right) =\int_{\mathcal{B}_{0}}\left( e_{E}\left( x 
\mathbf{,}u\left( x\right) ,F\left( x\right) \mathbf{,}\nu \left( x\right)
\right) +e_{M}\left( x\mathbf{,}\nu \left( x\right) \mathbf{,}N\left(
x\right) \right) \right) \text{ }dx.  \label{GL-ext}
\end{equation}

Special expressions of the decomposed energy density describe
ferroelectrics, spin glasses, liquid crystals, affine bodies etc. More
specifically, the density in (\ref{GL-ext}) is a generalized form of the
Ginzburg-Landau energy 
\begin{equation*}
e_{E}\left( x,\nu \right) +\frac{1}{2}\varpi \left\vert N\right\vert ^{2},
\end{equation*}%
with $e_{E}\left( x,\nu \right) $ a non-homogeneous two-well energy and $%
\varpi $\ an appropriate material constant.

From a physical point of view a constitutive choice of the type (\ref{GL-ext}
) is like to imagine that the substructural action due to the relative
change of the substructural shapes between neighboring material elements (an
action called \emph{microstress}) is not influenced directly by the
macroscopic deformation, the interplay being only indirect.

For example, take into account a body in which $\nu \left( x\right) $ is a
second rank symmetric tensor with components $\nu _{\alpha \beta }$, a
tensor measuring a micro-deformation of each material element, a deformation
independent of the macroscopic one, as it occurs in soft bodies with
families of polymeric chains scattered in a melt (see e.g. \cite{Mn}). The
manifold of substructual shapes $\mathcal{M}$ then coincides with the linear
space of second rank symmetric tensors over $\mathbb{R}^{3}$. For the sake
of simplicity, one may also make use here of the displacement vector $%
\mathsf{u}=i^{-1}\left( u\left( x\right) \right) -x$, with $i$ the
isomorphism between $\mathbb{R}^{3}$ and $\mathbb{\hat{R}}^{3}$. The elastic
energy depends on $\left( D\mathsf{u}\right) _{ij}$, $\nu _{\alpha \beta }$
and $N_{\alpha \beta i}$. Take note that in this example there is no
distinction between covariant and contravariant components for the sake of
simplicity. Latin indices indicate coordinates in the ambient space while
Greek indices label coordinates over $\mathcal{M}$. In infinitesimal
deformation and linear elastic setting, the elastic energy density takes the
form 
\begin{equation*}
e^{i}\left( D\mathsf{u},\nu ,N\right) =\frac{1}{2}C_{ijhk}\left( D\mathsf{u}%
\right) _{ij}\left( D\mathsf{u}\right) _{hk}+
\end{equation*}%
\begin{equation*}
+A_{ij\alpha \beta }^{1}\left( D\mathsf{u}\right) _{ij}\nu _{\alpha \beta
}+A_{ij\alpha \beta k}^{2}\left( D\mathsf{u}\right) _{ij}N_{\alpha \beta k}+
\end{equation*}%
\begin{equation*}
+\frac{1}{2}A_{\alpha \beta \gamma \delta }^{3}\nu _{\alpha \beta }\nu
_{\gamma \delta }+A_{\alpha \beta \gamma \delta k}^{4}\nu _{\alpha \beta
}N_{\alpha \beta k}+\frac{1}{2}A_{\alpha \beta i\gamma \delta
j}^{5}N_{\alpha \beta i}N_{\gamma \delta j},
\end{equation*}%
with the constitutive tensors $C$ and $A^{k}$, $k=1,\dots ,5$, endowed at
least with `major' symmetries. If the material texture is centrosymmetric, a
property that can be assumed commonly for each material element, by standard
group calculations one may verify that all odd constitutive tensors vanish,
namely $A_{ij\alpha \beta k}^{2}=0,$ \ \ $A_{\alpha \beta \gamma \delta
k}^{4}=0$, so that the energy reduces to a very special case of (\ref{GL-ext}%
), precisely 
\begin{equation*}
e^{i}\left( D\mathsf{u},\nu ,D\nu \right) =\frac{1}{2}C_{ijhk}\left( D%
\mathsf{u}\right) _{ij}\left( D\mathsf{u}\right) _{hk}+A_{ij\alpha \beta
}^{1}\left( D\mathsf{u}\right) _{ij}\nu _{\alpha \beta }+
\end{equation*}%
\begin{equation*}
+\frac{1}{2}A_{\alpha \beta \gamma \delta }^{3}\nu _{\alpha \beta }\nu
_{\gamma \delta }+\frac{1}{2}A_{\alpha \beta i\gamma \delta j}^{5}N_{\alpha
\beta i}N_{\gamma \delta j}.
\end{equation*}

Such a reduction does not occur when, for example, the morphological
descriptor is a vector with generic component $\nu _{\alpha }$, a vector
belonging to some copy of $\mathbb{R}^{3}$. In infinitesimal deformation and
linear elastic setting, the elastic energy density takes in this case the
form 
\begin{equation*}
e^{i}\left( D\mathsf{u},\nu ,D\nu \right) =\frac{1}{2}C_{ijhk}\left( D%
\mathsf{u}\right) _{ij}\left( D\mathsf{u}\right) _{hk}+
\end{equation*}%
\begin{equation*}
+A_{ij\alpha }^{1}\left( D\mathsf{u}\right) _{ij}\nu _{\alpha }+A_{ij\alpha
k}^{2}\left( D\mathsf{u}\right) _{ij}N_{\alpha k}+
\end{equation*}%
\begin{equation*}
+\frac{1}{2}A_{\alpha \gamma }^{3}\nu _{\alpha }\nu _{\gamma }+A_{\alpha
\gamma k}^{4}\nu _{\alpha }N_{\gamma k}+\frac{1}{2}A_{\alpha i\gamma
j}^{5}N_{\alpha i}N_{\gamma j},
\end{equation*}%
with the constitutive tensors $C$ and $A^{k}$, $k=1,\dots ,5$, endowed with
`major' symmetries. If the material is centrosymmetric, then the expression
of the energy reduces to 
\begin{equation*}
e^{i}\left( D\mathsf{u},\nu ,D\nu \right) =\frac{1}{2}C_{ijhk}\left( D%
\mathsf{u}\right) _{ij}\left( D\mathsf{u}\right) _{hk}+A_{ij\alpha
k}^{2}\left( D\mathsf{u}\right) _{ij}N_{\alpha k}+
\end{equation*}%
\begin{equation*}
+\frac{1}{2}A_{\alpha \gamma }^{3}\nu _{\alpha }\nu _{\gamma }+\frac{1}{2}%
A_{\alpha i\gamma j}^{5}N_{\alpha i}N_{\gamma j}.
\end{equation*}%
This energy density appears in the mechanics of microcracked bodies (see 
\cite{MaSt05}) and in the one of quasi-periodic alloys (see \cite{M06}) in
infinitesimal deformation regime and is not a special case of \eqref{Gen}. A
special expression of the energuy above is adeguate for the mechanics of
quasicrystals.

\section{Ground states: existence theorems}

Two steps are in general necessary for determining existence of minimizers
of functionals by the direct methods of calculus of variations: (\emph{i})
the extension of the functional class of competitors to some topological
space in such a way that energy bounded sets are compact, and (\emph{ii})
the appropriate extension of the energy functional over this enlarged class
as a lower semicontinuous function. The first requirement forces to
introduce a class of competitors which includes non-smooth functions. Once
the class of competitors is selected, the extended energy under scrutiny is
defined as the relaxed energy. However, even in the setting of finite
elasticity of simple bodies, the relaxed energy is not known actually, so
that only heuristic choices for (\emph{i}) and (\emph{ii}) can be made.

In the more general context of complex bodies considered here, a similar
procedure is followed.

The results do not exclude apriori a Lavrentiev gap phenomenon. Moreover,
with respect to the non-linear elasticity of simple bodies, the situation
tackled is even more intricate for the presence of the interplay between
gross and substructural changes.

Take note that the choice of the topology and of the functional extensions
mentioned above requires additional assumptions on the structure of the
energy and the functional class of competitors. Such assumptions have often
a non-trivial physical meaning (a meaning that has to be clarified at least
in special cases of prominent physical interest) and, in this sense, they
have constitutive nature. The effect of such assumptions becomes prominent
when they imply the existence of singular minimizers and/or phenomena of
localization of energy.

\subsection{Preliminaries}

Let $I\left( k,n\right) $ be the space of multi-indices of length $k$.
Denote also by $0$ the empty multi-index of length $0$. For any $\alpha $,
the\emph{\ complementary }multi-index to $\alpha $ in $\left(
1,2,...,n\right) $ is denoted by $\bar{\alpha}$ and the sign of the
permutation from $\left( 1,2,...,n\right) $ into $\left( \alpha
_{1},...,\alpha _{k},\bar{\alpha}_{1},...,\bar{\alpha}_{n-k}\right) $ is
indicated by $\sigma \left( \alpha ,\bar{\alpha}\right) $ .

For $\left( \mathbf{e}_{1},...,\mathbf{e}_{n}\right) $ and $\left( \mathbf{\
\varepsilon }_{1},...,\mathbf{\varepsilon }_{N}\right) $ bases in $\mathbb{R}%
^{n}$ and $\mathbb{R}^{N}$ respectively, $\Lambda _{r}\left( \mathbb{R}%
^{n}\times \mathbb{R}^{N}\right) $ indicates the vector space of
skew-symmetric tensors over $\mathbb{R}^{n}\times \mathbb{R}^{N}$ of the
form 
\begin{equation*}
\xi =\sum_{\left\vert \alpha \right\vert +\left\vert \beta \right\vert
=r}\xi ^{\alpha \beta }\mathbf{e}_{\alpha }\wedge \mathbf{\varepsilon }%
_{\beta }=\sum_{\max \left( 0,r-n\right) }^{\min \left( r,N\right) }\xi
_{\left( k\right) },
\end{equation*}%
where 
\begin{equation*}
\xi _{\left( k\right) }=\sum_{\substack{ \left\vert \alpha \right\vert
+\left\vert \beta \right\vert =r  \\ \left\vert \beta \right\vert =k}}\xi
^{\alpha \beta }\mathbf{e}_{\alpha }\wedge \mathbf{\varepsilon }_{\beta }.
\end{equation*}%
The decomposition $\xi =\sum_{k}\xi _{\left( k\right) }$ does not depend on
the choice of the bases.

For any linear map $G:\mathbb{R}^{n}\rightarrow \mathbb{R}^{N}$, the
notation $M\left( G\right) $ is used for the simple $n-$vector in $\Lambda
_{n}\left( \mathbb{R}^{n}\times \mathbb{R}^{N}\right) $ tangent to the graph
of $G$ and defined by 
\begin{eqnarray*}
M\left( G\right) &:&=\Lambda _{n}\left( id\times G\right) \left( \mathbf{e}%
_{1}\wedge ...\wedge \mathbf{e}_{n}\right) \\
&=&\left( \mathbf{e}_{1},G\left( \mathbf{e}_{1}\right) \right) \wedge
...\wedge \left( \mathbf{e}_{n},G\left( \mathbf{e}_{n}\right) \right) .
\end{eqnarray*}%
In coordinates one gets 
\begin{equation*}
M\left( G\right) =\sum_{k=0}^{\min \left( n,N\right) }M_{\left( k\right)
}\left( G\right) ,
\end{equation*}%
where 
\begin{equation*}
M_{\left( k\right) }\left( G\right) =\sum_{\substack{ \left\vert \alpha
\right\vert +\left\vert \beta \right\vert =n  \\ \left\vert \beta
\right\vert =k}}\sigma \left( \alpha ,\bar{\alpha}\right) M_{\bar{\alpha}%
}^{\beta }\left( \mathbf{G}\right) \mathbf{e}_{\alpha }\wedge \mathbf{%
\varepsilon }_{\beta }.
\end{equation*}%
$\mathbf{G}$ indicates the matrix associated with $G$. Moreover, $M_{\bar{%
\alpha}}^{\beta }\left( \mathbf{G}\right) $ is the determinant of the
submatrix of $\mathbf{G}$ made of the rows and the columns indexed by $\beta 
$ and $\bar{\alpha}$ respectively. It is also convenient to put $%
M_{0}^{0}\left( \mathbf{G}\right) :=1$. In the special case in which $n=N=3$%
, the components of $M\left( G\right) $ are the entries of $G$, $%
\mathop{\rm
adj}\nolimits G$ and $\mathop{\rm det}\nolimits G.$

For $\mathcal{M}$ a smooth manifold, $\Lambda _{r}\left( \mathcal{M}\right) $%
\ can be also defined as $\Lambda _{r}\left( \mathcal{M}\right) :=\cup _{\nu
\in \mathcal{M}}\Lambda _{r}\left( T_{\nu }\mathcal{M}\right) $. The
definition is natural, because $T_{\nu }\mathcal{M}$ a linear space. Related
definitions can be then adapted. The natural vector algebra over the fiber $%
\Lambda _{r}\left( T_{\nu }\mathcal{M}\right) $ is extended over the fiber
bundle $\Lambda _{r}\left( T\mathcal{M}\right) $.

Consider the two orthogonal subspaces $\mathbb{R}^{3}$ and $\mathbb{R}^{N}$
of the Euclidean space $\mathbb{R}^{3}\times \mathbb{R}^{N}$, select $%
\mathcal{B}_{0}$ as a smooth open domain of $\mathbb{R}^{3}$ and take $u:%
\mathcal{B}_{0}\rightarrow $ $\mathbb{R}^{N}$.

For $u\in W^{1,1}\left( \mathcal{B}_{0},\mathbb{R}^{N}\right) $, let $%
\mathcal{\tilde{B}}_{0}$ be the subset of $\mathcal{B}_{0}$ of Lebesgue
points for both $u$ and $Du$. Let also $\tilde{u}$ be a Lusin representative
of $u$, and $\tilde{u}\left( x\right) $ and $D\tilde{u}\left( x\right) $ the
Lebesgue values of $u$ and $Du$\ at $x\in \mathcal{\tilde{B}}_{0}$. The
Lusin-type theorem for $W^{1,1}$ functions implies that the graph of $u$,
namely 
\begin{equation*}
\mathcal{G}_{u}:=\left\{ \left( x,y\right) \in \mathcal{B}_{0}\times \mathbb{%
\ R}^{N}\text{ }|\text{ }x\in \mathcal{\tilde{B}}_{0}\text{, }y=\tilde{u}
\left( x\right) \right\} ,
\end{equation*}
is a $3-$rectifiable subset of $\mathcal{B}_{0}\times \mathbb{R}^{N}$ with
approximate tangent vector space at $\left( x,u\left( x\right) \right) $
generated by the vectors $\left( \mathbf{e}_{1},Du\left( x\right) \mathbf{e}
_{1}\right) ,...,\left( \mathbf{e}_{n},Du\left( x\right) \mathbf{e}
_{n}\right) $.

For any $u\in W^{1,1}\left( \mathcal{B}_{0},\mathbb{R}^{N}\right) $ with $%
\left\vert M\left( Du\left( \mathbf{x}\right) \right) \right\vert \in
L^{1}\left( \mathcal{B}_{0}\right) $, the \emph{$3-$current integration}
over the graph of $u$ is the linear functional on smooth $3-$forms with
compact support in $\mathcal{B}_{0}\times \mathbb{R}^{N}$ defined by 
\begin{eqnarray*}
G_{u} &:&=\int_{\mathcal{B}_{0}}\left( id\times u\right) ^{\#}\left( \omega
\right) =\int_{\Omega }\left\langle \left( id\times u\right) ^{\#}\left(
\omega \right) ,\mathbf{e}_{1}\wedge ...\wedge \mathbf{e}_{n}\right\rangle 
\text{ }dx \\
&=&\int_{\mathcal{B}_{0}}\left\langle \omega \left( x,u\left( x\right)
\right) ,M\left( Du\left( x\right) \right) \right\rangle \text{ }dx,
\end{eqnarray*}%
where \# indicates pull-back of forms $\omega $. By the area formula 
\begin{equation}
G_{u}=\int_{\mathcal{G}_{u}}\left\langle \omega ,\xi \right\rangle \text{ }d%
\mathcal{H}^{3}\llcorner \mathcal{G}_{u}\mathbf{,}  \label{CurrSob}
\end{equation}%
where $\xi \left( x\right) :=\frac{M\left( Du\left( x\right) \right) }{%
\left\vert M\left( Du\left( x\right) \right) \right\vert }$ and $x\in \ 
\mathcal{B}_{0}$. $\xi $\ is the unit $3-$vector that orients the
approximate tangent plane to $\mathcal{G}_{u}$ at $x$; moreover, $G_{u}$ has
finite mass $\mathbf{M}\left( G_{u}\right) :=\sup_{\left\Vert \omega
\right\Vert _{\infty }\leq 1}G_{u}\left( \omega \right) $ since 
\begin{equation*}
\mathbf{M}\left( G_{u}\right) =\int_{\mathcal{B}_{0}}\left\vert M\left(
Du\left( x\right) \right) \right\vert \text{ }dx=\mathcal{H}^{3}\left( 
\mathcal{G}_{u}\right) .
\end{equation*}%
In particular, $G_{u}$ is a vector valued measure on $\mathcal{B}_{0}\times 
\mathbb{R}^{N}$. It is common usage to say that $G_{u}$ is an integer
rectifiable $3-$current with integer multiplicity 1 on $\mathcal{B}%
_{0}\times \mathbb{R}^{N}$.

For a generic function $u\in W^{1,1}\left( \mathcal{B}_{0},\mathbb{R}%
^{N}\right) $ with $\left\vert M\left( Du\left( x\right) \right) \right\vert
\in L^{1}\left( \mathcal{B}_{0}\right) $, in general the \emph{boundary
current} $\partial G_{u}$, defined by%
\begin{equation*}
\partial G_{u}\left( \omega \right) :=G_{u}\left( d\omega \right) ,\text{ \
\ \ }\omega \in \mathcal{D}^{2}\left( \mathcal{B}_{0}\times \mathbb{R}%
^{N}\right) ,
\end{equation*}
does not vanish, although $\partial G_{u}\left( \omega \right) =0$ for all $%
\omega \in \mathcal{D}^{2}\left( \mathcal{B}_{0}\times \mathbb{R}^{N}\right) 
$ if $u$ is smooth. A typical example is the map $u\left( x\right) :=\frac{x%
}{\left\vert x\right\vert }$ that belongs to $W^{1,2}\left( B^{3}\left(
0,1\right) ,\mathbb{R}^{3}\right) $. One computes that $\partial
G_{u}=-\delta _{0}\times S^{2}$ on $\mathcal{D}^{2}\left( B^{3}\left(
0,1\right) \times \mathbb{R}^{3}\right) $, with $\delta _{0}$ Dirac delta.
However, by approximating a map $u$ in the Sobolev norm by means of $C^{2}$
maps, it is easy to prove that $\partial G_{u}=0$ on $\mathcal{D}^{2}\left( 
\mathcal{B}_{0}\times \mathbb{R}^{N}\right) $ if $u\in W^{1,3}\left( 
\mathcal{B}_{0},\mathbb{R}^{N}\right) $.

\subsection{Functional characterization of the gross deformation}

The transplacement field is considered here as a weak diffeomorphism.

\begin{definition}
Let $u\in W^{1,1}(\mathcal{B}_{0},\mathbb{\hat{R}}^{3})$. $u$ is said a 
\emph{weak diffeomorphism} (one writes $u\in dif^{1,1}\left( \mathcal{B}_{0},%
\mathbb{R}^{3}\right) $) if

\begin{enumerate}
\item $\left\vert M\left( Du\left( x\right) \right) \right\vert \in
L^{1}\left( \mathcal{B}_{0}\right) ,$

\item $\partial G_{u}=0$ on $\mathcal{D}^{2}(\mathcal{B}_{0},\mathbb{\hat{R}}
^{3}),$

\item $\mathop{\rm det}\nolimits Du\left( x\right) >0$ a.e. $x\in \mathcal{B}%
_{0},$

\item for any $f\in C_{c}^{\infty }\left( \mathcal{B}_{0}\times \mathbb{R}
^{3}\right) $ 
\begin{equation}
\int_{\mathcal{B}_{0}}f\left( x\mathbf{,}u\left( x\right) \right) %
\mathop{\rm det}\nolimits Du\left( x\right) \text{ }dx\leq \int_{\mathbb{R}%
^{3}}\sup_{x\mathbf{\in } \mathcal{B}_{0}}f\left( x,y\right) dy\mathbf{.}
\label{GM}
\end{equation}
\end{enumerate}
\end{definition}

The requirement~3 above is the standard condition assuring that the map $%
\tilde{u}$ be orientation preserving. Condition~4 is a global one-to-one
condition, while items~1 and~2 provide the necessary uniformity. For
instance, the norm of the minors of $Du$, namely $\left\vert M\left(
Du\right) \right\vert $, is simply the standard square norm.

Condition 4 has been introduced in the form 
\begin{equation}
\int_{\mathcal{B}_{0}}\left( \mathop{\rm det}\nolimits Du\left( x\right)
\right) \text{ }dx\leq vol\left( u\left( \mathcal{B}_{0}\right) \right) .
\label{CN}
\end{equation}%
in \cite{CN} where the discussion is limited to the macroscopic deformation
of (simple) bodies in the setting of $W^{1,3}(\mathcal{B}_{0},\mathbb{\hat{R}%
}^{3})$ maps. It allows frictionless contact of parts of the boundary of the
body while still prevents the penetration of matter. Condition (\ref{CN}) is
equivalent to (\ref{GM}) with $u\left( \mathcal{B}_{0}\right) $ substituted
by $u\left( \mathcal{\tilde{B}}_{0}\right) $, with $\mathcal{\tilde{B}}_{0}$
the subset of $\mathcal{B}_{0}$ of Lebesgue points of $u$ and $Du$. The
version (\ref{GM}) is more useful for the analysis presented here.

Essential properties of weak diffeomorphisms are collected in the theorem
below, the proof of which can be found in \cite{GMS3}, where such a class of
diffeomorphisms has been introduced.

\begin{theorem}
\begin{enumerate}
\item {} (Closure) Let $\left\{ u_{k}\right\} $ be a sequence with $u_{k}\in
dif^{1,1}(\mathcal{B}_{0},\mathbb{\hat{R}}^{3})$ for any $k$. If 
\begin{equation*}
u_{k}\rightharpoonup u\text{ \ \ \ and \ \ \ }M\left( Du_{k}\right)
\rightharpoonup v
\end{equation*}
weakly in $L^{1}$, then $v=M\left( Du\right) $ a.e. and $u\in dif^{1,1}( 
\mathcal{B}_{0},\mathbb{\hat{R}}^{3})$.

\item {} (Compactness) Let $\left\{ u_{k}\right\} $ be a sequence with $%
u_{k}\in W^{1,r}(\mathcal{B}_{0},\mathbb{\hat{R}}^{3})$, $r>1$, and $u_{k}$
's weak diffeomorphisms. Assume that there exists a constant $C>0$ and a
convex function $\vartheta :\left[ 0,+\infty \right) \rightarrow \mathbb{R}%
^{+}$ such that $\vartheta \left( t\right) \rightarrow +\infty $ as $%
t\rightarrow 0^{+}$, and 
\begin{equation*}
\left\Vert M\left( Du_{k}\right) \right\Vert _{L^{r}\left( \mathcal{B}%
_{0}\right) }\leq C,\text{ \ \ \ \ }\int_{\mathcal{B}_{0}}\vartheta \left( %
\mathop{\rm det}\nolimits Du_{k}\left( x\right) \right) \text{ }dx\leq C.
\end{equation*}%
Then, by taking subsequences $\left\{ u_{j}\right\} $ with $%
u_{j}\rightharpoonup u$ in $W^{1,r}(\mathcal{B}_{0},\mathbb{\hat{R}}^{3})$,
one gets $u_{j}\rightarrow u$ in $L^{r}\left( \mathcal{B}_{0}\right) $, $%
M\left( Du_{j}\right) \rightharpoonup M\left( Du\right) $ in $L^{r}$ and $%
\int_{\mathcal{B}_{0}}\vartheta \left( \mathop{\rm det}\nolimits Du\left(
x\right) \right) $ $dx\leq C$. In particular, $u$ is a weak diffeomorphism.
\end{enumerate}
\end{theorem}

In particular, below it is assumed that the gross deformation is an element
of%
\begin{equation*}
dif^{r,1}(\mathcal{B}_{0},\mathbb{\hat{R}}^{3}):=\left\{ u\in dif^{1,1}(%
\mathcal{B}_{0},\mathbb{\hat{R}}^{3})|\left\vert M\left( Du\right)
\right\vert \in L^{r}\left( \mathcal{B}_{0}\right) \right\} ,
\end{equation*}%
for some $r>1$.

\subsection{Functional characterization of the morphological descriptor maps}

Above it has been emphasized that at each $x\in \mathcal{B}_{0}$ one gets $%
N\in Hom\left( T_{x}\mathcal{B}_{0},T_{\nu \left( x\right) }\mathcal{M}%
\right) \simeq \mathbb{R}^{3}\otimes T_{\nu }\mathcal{M}$. When the energy
density $e$ admits derivative with respect to $N$, such a derivative
describes the weakly non-local (gradient-conjugated) interactions between
neighboring material elements, interactions (the so-called microstresses)
due to relative changes in the substructural shapes. When one wants to
compute in covariant way $N$ explicitly, a connection over $\mathcal{M}$ is
necessary. Its choice determines the representation of the microstress and,
in this sense, it has constitutive nature. Physics may suggest also that a
connection on $\mathcal{M}$ has no physical meaning as in the case of
liquids with `dispersed' bubbles. Moreover, even when $\mathcal{M}$ is
selected to be Riemannian, in some circumstances the gauge needed for $N$
might not be the Levi-Civita one (see, e.g., \cite{CG}). In this case, if no
prevalent role is given to the Levi-Civita connection, the parallel
transport over geodetics may be non-isometric in general and also it can be
even unbounded as a consequence of topological features of $\mathcal{M}$
itself. The metric $g_{\mathcal{M}}$ generating the Levi-Civita connection
has also non-trivial physical meaning. In fact, as already mentioned in
Section~3, when the material substructure admits its own kinetic energy, its
first approximation is quadratic in the rates of the morphological
descriptors, a quadratic form with coefficients given by $g_{\mathcal{M}}$.
Conversely, if the quadratic substructural kinetic energy is prescribed by
experimental data, its coefficients determine the metric itself. In this
case, the related Levi-Civita connection brings information from
substructural kinetics. Consequently, if the covariant gradient of $\nu $\
is calculated by making use of the natural Levi-Civita connection, in
dynamic setting the substructural kinetics may directly determine the
representation of the microstress.

By Nash theorem, $\mathcal{M}$ is considered as a \emph{submanifold} in $%
\mathbb{R}^{N}$ for some appropriate dimension. In addition it is assumed
that $\mathcal{M}$ is closed. The covariant derivative of the map $\nu $ is
in agreement with the differential of $\nu $ as a map from $\mathcal{B}_{0}$
into $\mathbb{R}^{N}$.

With the premises above it is assumed that $\nu $ belongs to the Sobolev
space $W^{1,s}\left( \mathcal{B}_{0},\mathcal{M}\right) $, $s>1$, defined by%
\begin{equation*}
W^{1,s}\left( \mathcal{B}_{0},\mathcal{M}\right) :=\left\{ \nu \in
W^{1,s}\left( \mathcal{B}_{0},\mathbb{R}^{N}\right) \text{ }|\text{ }\nu
\left( x\right) \in \mathcal{M}\text{ for a.e. }x\in \mathcal{B}_{0}\right\}
.
\end{equation*}

Remarkably, since the isometric embedding of $\mathcal{M}$ in $\mathbb{R}%
^{N} $ by Nash's theorem is not unique, its choice (then the choice of the
appropriate space $W^{1,s}\left( \mathcal{B}_{0},\mathcal{M}\right) $) is an
additional constitutive prescription.

\subsection{Boundary conditions}

Boundary data have to be assigned for both the fields $u$ and $\nu $. As
regards the macroscopic deformation $u$, place, traction or mixed data can
be prescribed. As regards Dirichlet data, the ones considered here, since
the deformation $u$ is assumed to be an element of a Sobolev space, its
value along the boundary $\partial \mathcal{B}_{0}$ or along an open part $%
\partial \mathcal{B}_{0,u}$ of it has to be assigned in the sense of traces.
If $Du$ admits summable minors, there is also the possibility to fix the
value of the current $\partial G_{u}$ as a functional on $\mathcal{D}^{2}(%
\mathbb{R}^{3}\times \mathbb{\hat{R}}^{3})$, a condition called \emph{strong
anchoring}. Notice that $\partial G_{u_{k}}\rightharpoonup \partial G_{u}$
if $M\left( Du_{k}\right) \rightarrow M\left( Du\right) $ in $L^{1}$. The
assignment of $\partial G_{u}$ is, in fact, stronger than prescribing the
sole trace of $u$ and reduces to it if $u\in W^{1,r}(\mathcal{B}_{0},\mathbb{%
\hat{R}}^{3})$, $r\geq 3$, or $u\in W^{1,2}(\mathcal{B}_{0},\mathbb{\hat{R}}%
^{3})$ and $\mathop{\rm adj}\nolimits Du\in L^{\frac{3}{2}}$.

Boundary data of Dirichlet type are naturally admissible also for $\nu $ and
they should be intended in the sense of traces. In fact, in some physical
circumstances one may prescribe the shape of the substructure at the
boundary of a complex body. The prototype example is the one of liquid
crystals in nematic phase. In this case, the substructure is made of stick
molecules with end-to-tail symmetry embedded in a ground liquid. For
example, in a channel the orientation of the stick molecules along the walls
of the channel can be prescribed by means of the use of surfactants spread
along the walls themselves (see, e.g., \cite{CB}). However, when shrewdness
of this type are not available, a natural choice is to imagine that the
material elements at the boundary do not undergo substructural changes so
that one may prescribe that $\nu $ vanishes identically if the null value is
included in $\mathcal{M}$.

In special circumstances one may think of the material elements on the
boundary as made of simple material and that the complexity of the matter
vanishes in a boundary layer: the relevant theory is not yet developed.

Moreover, one could consider the boundary as a sort of membrane coating the
body. This point of view is helpful when one would like to assign data in
terms of substructural tractions given by the conormal derivative $\partial
_{D\nu }e$ $n$, with $n$ the normal to the boundary in the points in which
it is defined. In fact, devices able to assign along the external boundary
of the body contact direct actions on the substructure inside each material
element seem to be not available, so that the natural boundary condition for
substructural tractions is $\partial _{D\nu }e$ $n=0$. Such a condition
makes always sense because at each $x$ the conormal derivative $\partial
_{D\nu }e$ $n$ is an element of the cotangent space $T_{\nu \left( x\right)
}^{\ast }\mathcal{M}$, a linear space indeed, containing the null value. The
external boundary of the body can be also considered as a structured surface
endowed with a surface energy depending on the normal (if the boundary is
anisotropic), the surface gradient of deformation, the curvature tensor, the
morphological descriptor and its surface gradient. In this case, the
derivative of the surface energy with respect to the gradient of the
morphological descriptor, applied to the normal, is the boundary datum in
term of substructural action \cite{M06}. Of course, the choice of an
explicit expression of the surface energy is strictly of constitutive nature.

\subsection{Existence results}

The portions $\partial \mathcal{B}_{0,u}$ and $\partial \mathcal{B}_{0,\nu }$
of the boundary where Dirichlet data are prescribed, may or may not coincide
with the whole $\partial \mathcal{B}_{0}$. On $\partial \mathcal{B}%
_{0}\backslash \partial \mathcal{B}_{0,u}$ and $\partial \mathcal{B}%
_{0}\backslash \partial \mathcal{B}_{0,\nu }$ it is assumend that
macroscopic and microscopic tractions satisfy the natural homogeneous null
condition.

Define the space $\mathcal{W}_{r,s}$ by 
\begin{equation*}
\mathcal{W}_{r,s}:=\left\{ \left( u,\nu \right) |u\in dif^{r,1}(\mathcal{B}
_{0},\mathbb{\hat{R}}^{3}),\text{ }\nu \in W^{1,s}\left( \mathcal{B}_{0}, 
\mathcal{M}\right) \right\} .
\end{equation*}

Imagine also that the energy functional (\ref{Gen}) is extended to $\mathcal{%
\ W}_{r,s}$ as 
\begin{equation*}
\mathcal{E}\left( u,\nu \right) =\int_{\mathcal{B}_{0}}e\left( x\mathbf{,}
u\left( x\right) ,Du\left( x\right) \mathbf{,}\nu \left( x\right) \mathbf{,}
D\nu \left( x\right) \right) \text{ }dx\mathbf{,}
\end{equation*}
where $u\left( x\right) $, $Du\left( x\right) $, $\nu \left( x\right) $ and $%
D\nu \left( x\right) $ are the Lebesgue values of $u$, $\nu $ and their weak
derivatives.

Constitutive assumptions about the structure of the energy density $e$ are
also necessary. They are additional to the ones described in the previous
sections concerning the functional nature of the fields involved.

Consider the energy density $e$ as a map 
\begin{equation*}
e:\mathcal{B}_{0}\times \mathbb{\hat{R}}^{3}\times \mathcal{M}\times
M_{3\times 3}^{+}\times M_{N\times 3}\rightarrow \mathbb{\bar{R}}^{+}
\end{equation*}%
with values $e\left( x,u,F\mathbf{,}\nu ,N\right) $. The assumptions below
about $e$ apply.

\begin{description}
\item[(H1)] $e$ is polyconvex in $F$ and convex in $N$. More precisely,
there exists a Borel function 
\begin{equation*}
Pe:\mathcal{B}_{0}\times \mathbb{\hat{R}}^{3}\times \mathcal{M}\times
\Lambda _{3}(\mathbb{R}^{3}\times \mathbb{\hat{R}}^{3})\times M_{N\times
3}\rightarrow \mathbb{\bar{R}}^{+},
\end{equation*}
with values $Pe\left( x,u,\nu ,\xi ,N\right) $, which is

\begin{description}
\item {\emph{(a)}} l. s. c. in $\left( u,\nu ,\xi ,N\right) $ for a.e. $x\in 
\mathcal{B}_{0}$,

{\emph{(b)}} convex in $\left( \xi ,N\right) $\ for any $\left( x,u,\nu
\right) $,

{\emph{(c)}} such that $Pe\left( x,u,\nu ,M\left( F\right) \mathbf{,}%
N\right) =e\left( x,u,\nu ,F\mathbf{,}N\right) $ for any list of entries $%
\left( x,u,\nu ,F\mathbf{,}N\right) $ with $\mathop{\rm det}\nolimits F>0$.
\end{description}

In terms of $Pe$, the energy functional becomes 
\begin{equation}
\mathcal{E}\left( u,\nu \right) =\int_{\mathcal{B}_{0}}Pe\left( x\mathbf{,}
u\left( x\right) ,\nu \left( x\right) \mathbf{,}M\left( F\right) \mathbf{,}
N\right) \text{ }dx\mathbf{.}  \label{Policon}
\end{equation}

\item[(H2)] The energy density $e$ satisfies the growth condition 
\begin{equation}
e\left( x,u,\nu ,F\mathbf{,}N\right) \geq C_{1}\left( \left\vert M\left(
F\right) \right\vert ^{r}+\left\vert N\right\vert ^{s}\right) +\vartheta
\left( \mathop{\rm det}\nolimits F\right)  \label{Growth}
\end{equation}
for any $\left( x,u,\nu ,F\mathbf{,}N\right) $ with $\mathop{\rm det}%
\nolimits F>0$, $r,s>1$, $C_{1}>0$ constants and $\vartheta :\left(
0,+\infty \right) \rightarrow \mathbb{R}^{+}$ a convex function such that $%
\vartheta \left( t\right) \rightarrow +\infty $ as $t\rightarrow 0^{+}$.
\end{description}

The convexity of $Pe$ in $\left( \xi ,N\right) $\ for any $\left( x,y,\nu
\right) $ is in essence an assumption of stability which is more subtle than
usual. In fact, in standard elasticity the condition involves only $M\left(
F\right) $. Here there is an interplay between the gross deformation and the
substructure: the former must contribute to the stability of the latter and
vice versa.

The growth condition imposes that the set of admissible energies has a lower
bound which is a decomposed energy of Ginzburg-Landau type that generates
only interactions between neighboring material elements. Such actions are of
gradient type, and generate the so-called microstress. In other words, the
assumption (\ref{Growth}) means that, in a conservative setting,
substructural events within the generic material element, events that
generate self-actions, may only increase the energy.

If there is a pair $\left( u_{0},\nu _{0}\right) \in \mathcal{W}_{r,s}$ such
that $\mathcal{E}\left( u_{0},\nu _{0}\right) <+\infty $, from the closure
theorem for weak diffeomorphisms above and Ioffe's classical semicontinuity
result, the following theorem holds:

\begin{theorem}
Under the hypotheses (H1) and (H2) the functional $\mathcal{E}$ achieves the
minimum value in the classes 
\begin{equation*}
\mathcal{W}_{r,s}^{d}:=\left\{ \left( u,\nu \right) \in \mathcal{W}
_{r,s}|u=u_{0}\text{ on }\partial \mathcal{B}_{0,u},\nu =\nu _{0}\text{ on }
\partial \mathcal{B}_{0,\nu }\right\}
\end{equation*}
and 
\begin{equation*}
\mathcal{W}_{r,s}^{c}:=\left\{ \left( u,\nu \right) \in \mathcal{W}_{r,s} 
\text{ }|\text{ }\partial G_{u}=\partial G_{u_{0}}\text{ on }\mathcal{D}%
^{2}( \mathbb{R}^{3}\times \mathbb{\hat{R}}^{3}),\nu =\nu _{0}\text{ on }%
\partial \mathcal{B}_{0,\nu }\right\} .
\end{equation*}
\end{theorem}

This theorem extends traditional existence results for simple elastic bodies
and also the results for the minimizers of material substructures existing
in special cases when gross deformations are neglected. Moreover, it
indicates a path to characterize ground states in classes of bodies that
have been not investigated so far.

Several variants are possible.

\begin{description}
\item[(H3)] Assume that the energy density satisfies the growth condition 
\begin{equation}
e\left( x,u,\nu \mathbf{,}F\mathbf{,}N\right) \geq C_{2}(\left\vert
F\right\vert ^{2}+\left\vert Adj\text{ }F\right\vert ^{3/2}+\left\vert
N\right\vert ^{s})+\vartheta \left( \mathop{\rm det}\nolimits F\right)
\label{Gr}
\end{equation}%
for any $\left( x,u,\nu ,F,N\right) $ with $\mathop{\rm det}\nolimits F>0$, $%
C_{2}>0$ a constant and $\vartheta :\left( 0,+\infty \right) \rightarrow 
\mathbb{R}^{+}$ as above.
\end{description}

The growth condition (\ref{Gr}) has the same physical meaning of (\ref%
{Growth}), differences relying only in the explicit dependence of the lower
bound on the macroscopic deformation.

Define now the class $\mathcal{W}_{2,\frac{3}{2},s}$ as 
\begin{eqnarray*}
\mathcal{W}_{2,\frac{3}{2},s} &:&=\left\{ \left( u,\nu \right) |u\in
W^{1,2}( \mathcal{B}_{0},\mathbb{\hat{R}}^{3}),Adj(Du)\in L^{3/2},\right. \\
&&\left. \text{(4.) in Def. 1 holds},\nu \in W^{1,s}\left( \mathcal{B}_{0}, 
\mathcal{M}\right) \right\} .
\end{eqnarray*}

If the energy functional $\mathcal{E}$ is defined now on the class $\mathcal{%
W}_{2,\frac{3}{2},s}$, and there is a pair $\left( u_{0},\nu _{0}\right) \in 
\mathcal{W}_{2,\frac{3}{2},s}$ such that $\mathcal{E}\left( u_{0},\nu
_{0}\right) <+\infty $, on account of the $L\log L$ estimate in \cite{MTY},
the new existence result below follows.

\begin{theorem}
Under assumptions (H1) and (H3), the functional $\mathcal{E}$ achieves its
minimum value in the class 
\begin{equation*}
\mathcal{W}_{2,\frac{3}{2},s}^{d}:=\left\{ \left( u,\nu \right) \in \mathcal{%
\ W}_{2,\frac{3}{2},s}\text{ }|\text{ }u=u_{0}\text{ on }\partial \mathcal{B}
_{0,u},\nu =\nu _{0}\text{ on }\partial \mathcal{B}_{0,\nu }\right\} .
\end{equation*}
\end{theorem}

The special case of partially decomposed free energies (\ref{GL-ext}) falls,
of course, within the theorems above. Moreover, in the setting justifying
Theorem 2, the additive decomposition of the energy density $e$ in its \emph{%
\ macroscopic} and \emph{microscopic} parts $e_{E}$ and $e_{M}$ allows one
to separate the growth condition (\ref{Growth}) in two parts, namely 
\begin{equation}
e_{E}\left( x,u,\nu ,F\right) \geq C_{1}\left\vert M\left( F\right)
\right\vert ^{r}+\vartheta \left( \mathop{\rm det}\nolimits F\right) ,
\label{EE}
\end{equation}
\begin{equation}
e_{M}\left( x,u,\nu ,N\right) \geq C_{1}\left\vert N\right\vert ^{s}.
\label{EM}
\end{equation}

By fixing $\nu $ in (\ref{EE}), one recovers a growth condition rather
standard in finite elasticity of simple bodies where it is imposed only that 
$e_{E}\left( x,u,F\right) \geq C_{1}\left\vert M\left( F\right) \right\vert
^{r}+\vartheta \left( \mathop{\rm det}\nolimits F\right) $ (see, e.g., \cite%
{Sil}). This last requirement is tantamount to affirm that one is able to
find ground states for bodies with a content of energy greater or equal to
the one of a `fictitious' elastic simple body with energy given by $%
C_{1}\left\vert M\left( F\right) \right\vert ^{r}+\vartheta \left( %
\mathop{\rm det}\nolimits F\right) $. In using (\ref{EE}), however, one is
saying something more because of the presence of the morphological
descriptor $\nu $. With (\ref{EE}) it is prescribed that the standard lower
bound for simple bodies be also valid for the macroscopic part of the energy
of complex bodies admitting partially decomposed structure. The presumption
is that substructural events accruing within each material element do not
alter the lower bound, roughly speaking, substructural changes within the
material element may only increase the global energy in conservative
setting, at least with respect to $C_{1}\left\vert M\left( F\right)
\right\vert ^{r}+\vartheta \left( \mathop{\rm det}\nolimits F\right) $. No
matter about weakly non-local substructural interactions measured by the
microstress, namely by the derivative of the energy with respect to $D\nu $.
The energetics of such interactions is described by $e_{M}$. The condition (%
\ref{EM}) indicates that the energy accounting for both substructural
changes and weakly non-local interactions of gradient type admits as lower
bound the energy of a `fictitious' rigid complex material for which the
energy stored within each material element is negligible with respect to the
one associated with weakly non-local interactions. Specifically, the energy
of such a `fictitious' complex material is an extension of the Dirichlet
energy and reduces to it when $s=2$. In conservative case, the requirement (%
\ref{EM}) is quite natural because substructural activity within the
material element may only increase the energy density, being the energy
density of all events non-negative.

Analogous physical interpretations hold in the setting justifying Theorem 3
where (\ref{EE}) and (\ref{EM}) become respectively 
\begin{equation*}
e\left( x,u,\nu ,F\right) \geq C_{2}(\left\vert F\right\vert ^{2}+\left\vert
Adj\text{ }F\right\vert ^{2})+\vartheta \left( \mathop{\rm det}\nolimits
F\right) ,
\end{equation*}
\begin{equation*}
e_{M}\left( x,u,\nu ,N\right) \geq C_{2}\left\vert N\right\vert ^{2},
\end{equation*}
being $e_{M}$ in this case a Dirichlet energy when $C_{2}=\frac{1}{2}$.

\subsection{Remarks about the possible presence of a Lavrentiev gap
phenomenon}

There is no evidence that the energy functional 
\begin{equation*}
\mathcal{E}\left( u,\nu \right) =\int_{\mathcal{B}_{0}}Pe\left( x\mathbf{,}%
u\left( x\right) ,\nu \left( x\right) \mathbf{,}M\left( F\right) \mathbf{,}%
N\right) \text{ }dx
\end{equation*}%
defined over the class $\mathcal{W}_{r,s}$ is the relaxed version of the
same functional defined on regular pairs $\left( u,\nu \right) $, namely
there is no evidence that 
\begin{equation*}
\mathcal{E}\left( u,\nu \right) :=\inf \left\{ \liminf_{j\rightarrow \infty }%
\mathcal{E}\left( u_{j},\nu _{j}\right) \text{ }|\text{ }\left( u_{j},\nu
_{j}\right) \rightharpoonup \left( u,\nu \right) \text{ in }L^{1},\left(
u_{j},\nu _{j}\right) \in C^{1}\right\}
\end{equation*}%
for any $\left( u,\nu \right) \in \mathcal{W}_{r,s}$. In other words, a
Lavrentiev gap phenomenon, namely 
\begin{equation*}
\mathcal{E}\left( u,\nu \right) <\inf \left\{ \liminf_{j\rightarrow \infty }%
\mathcal{E}\left( u_{j},\nu _{j}\right) \text{ }|\text{ }\left( u_{j},\nu
_{j}\right) \rightharpoonup \left( u,\nu \right) \text{ in }L^{1},\left(
u_{j},\nu _{j}\right) \in C^{1}\right\} ,
\end{equation*}%
is not excluded a priori. Examples of special cases of the one treated here
are known from the scientific literature. For instance, in two-dimensional
ambient space there are examples of non-linear elastic simple materials
admitting a gap between the infimum of the energy over admissible continuous
deformations belonging to a Sobolev space $W^{1,r}$ and the analogous
infimum over admissible continuous deformations belonging to a Sobolev space 
$W^{1,s}$ with $s<r$ \cite{FHM}.

A gap phenomenon driven this time by the topology of the substructural
manifold $\mathcal{M}$ can also arise. When the manifold of substructural
shapes $\mathcal{M}$ has a non trivial homology, then defects may arise.

In particular, for example, a gap phenomenon appears for the Dirichlet
integral 
\begin{equation}
\mathcal{D}\left( \nu \right) :=\frac{1}{2}\int_{\mathcal{B}_{0}}\left\vert
D\nu \left( x\right) \right\vert ^{2}\text{ }dx  \label{DirIntegr}
\end{equation}%
involving maps $\nu :\mathcal{B}_{0}\rightarrow S^{2}$, with $S^{2}$ the
unit sphere, and for Dirichlet boundary data (see results in \cite{GM}, \cite%
{GMS1}, \cite{GMS2}). Take note that the exponent $2$ and the circumstance
that the manifold of substructural shapes is $S^{2}$ is crucial for the
remarks in what follows. The energy above, in absence of gross deformations,
describes cases of spin glasses, magnetostrictive materials in conditions of
magnetic saturation and soft composites reinforced with a dense family of
microfibers not endowed with end-to-tail symmetry.

The gap phenomenon appears when the degree of the boundary datum is
different from zero (because regular maps satisfying it are absent) and even
for some boundary data with zero degree \cite{HL}. However, it is possible
to find an explicit form for the relaxed version of (\ref{DirIntegr}) on $%
W^{1,2}$ which gives rise to a non-local functional \cite{BBC}.

A further difference between the maps in $W^{1,2}$ and the regular ones
relies in the behavior of the current 
\begin{equation*}
\mathbf{D}_{\nu }\left( \eta \right) :=\int_{\mathcal{B}_{0}}\eta \wedge \nu
^{\#}\omega _{S^{2}},\text{ \ \ \ \ \ }\eta \in \mathcal{D}^{1}\left( 
\mathcal{B}_{0}\right) ,
\end{equation*}
with $\omega _{S^{2}}$ the volume form over $S^{2}$, a current which can be
considered as integration along the field $D\left( x\right) $ defined by
duality by $D_{\nu }\left( x\right) :=\ast \omega _{S^{2}}$, with $\ast $
the Hodge star operator. Take note that, whereas $D_{\nu }\left( x\right)
\neq 0$, $D_{\nu }\left( x\right) $ generates $\mathop{\rm ker}\nolimits
D\nu \left( x\right) $ at $x$. Since the outward flux across the boundary $%
\partial B\left( x_{0},r\right) $ of a ball $B\left( x_{0},r\right) $,
centered at $x_{0}$ and with radius $r$, is the degree of the map $\nu
\left\vert _{B\left( \mathbf{x}_{0},r\right) }\right. $, it follows that $%
DivD_{\nu }=0$ for any $\nu \in C^{1}\left( \mathcal{B}_{0}\right) $. In
contrast, and for the same reason, for the map $\nu \left( x\right) :=\frac{x%
}{\left\vert x\right\vert } \in W^{1,2}\left( \mathcal{B}_{0},S^{2}\right) $
one gets $DivD_{\nu }=4\pi \delta _{0}$ in distributional sense.

By following results in \cite{GMS6}, \cite{GMS7}, \cite{GMu}, a way to link
the loss of energy in the Dirichlet integral and the `bad' behavior of some
functions in $W^{1,2}$ in pulling-back $2-$forms, is to associate to each
map $\nu \in W^{1,2}\left( \mathcal{B}_{0},S^{2}\right) $ the current
integration over its graph 
\begin{equation*}
G_{\nu }\left( \omega \right) =\int_{\mathcal{B}_{0}}\omega \left( x,\nu
\left( x\right) ,M\left( D\nu \left( x\right) \right) \right) \text{ }dx%
\mathbf{.}
\end{equation*}

In particular, it is proved that if $\left\{ \nu _{j}\right\} $ is a
sequence of $S^{2}$-valued maps with equibounded Dirichlet energies and $T$
is a current such that $G_{\nu _{j}}\rightharpoonup T$, then $T$ has finite
mass and there exists a map $\nu _{T}\in W^{1,2}\left( \mathcal{B}
_{0},S^{2}\right) $ and a one-dimensional integer rectifiable current $L_{T}$
, both map and current individuated uniquely by $T$, such that 
\begin{equation*}
T=G_{\nu _{T}}+L_{T}\times S^{2}\text{ on }\mathcal{D}^{3}\left( \mathcal{B}
_{0}\times S^{2}\right) .
\end{equation*}
Moreover, $L_{T}=0$ if $M\left( D\nu _{j}\right) $ weakly converges in $%
L^{1}(\mathcal{B}_{0},\Lambda _{3}(\mathbb{R}^{3},\mathbb{\hat{R}}^{3}))$.

The meaning of the concentration line $L_{T}$ is clear from the point of
view of weak convergence. In fact, if $G_{\nu _{j}}\rightharpoonup T=G_{\nu
_{T}}+L_{T}\times \mathcal{M}$, then on two-dimensional sections orthogonal
to the support of $L_{T}$ in a thin tube wrapped around the support itself,
for $j$ sufficiently large, $\nu _{j}$ assumes as value the entire sphere,
point by point. $L_{T}$ is then a line (a line defect) in which there is
`fusion' of the substructure, fusion in the sense of complete disorder so
that the concept of prevailing direction loses its meaning. Moreover, the
currents $D_{\nu _{j}}$, associated with the elements of the sequence $%
\left\{ \nu _{j}\right\} $, converge (in the sense of currents) to the
current $D_{T}=D_{\nu _{T}}+4\pi L_{T}$; in particular $\partial D_{T}=0$ on 
$\mathcal{B}_{0}\times S^{2}$. If $L_{T}=0$, the graph of $\nu _{T}$ has no
boundary and $DivD_{\nu _{T}}=0$. In this case, if $\nu _{T}$ would have
point singularities, all singularities would have zero degree.

It could be possible to extend the previous discussion by substituting $%
S^{2} $ with a generic two dimensional compact manifold $\mathcal{M}$, but $%
L_{T}=0$ if $\mathcal{M}$ is not homologically a sphere.

\subsection{Cartesian currents and the special case of spin substructures}

Dirichlet energies involving $S^{2}$-valued maps describe essential aspects
of the mechanical behavior of bodies with spin structure that does not
suffer deformation. Basic results have been obtained in the current
literature. Essential aspect are reviewed here first, then it is shown how
they can be extended to energies more general that the Dirichlet one, that
is to the description of the mechanical behavior od deformable bodies with
spin substructure. Basically the analysis is essentially the same, the
physics of the phenomena covered is enlarged drastically.

A current $T\in \mathcal{D}_{3}(\mathcal{B}_{0}\times \mathbb{\hat{R}}^{3})$
is in $cart^{2,1}\left( \mathcal{B}_{0}\times S^{2}\right) $ if there exist
a map $\nu _{T}\in W^{1,2}\left( \mathcal{B}_{0},S^{2}\right) $ and an
integer rectifiable current $L_{T}$ on $\mathcal{D}^{1}\left( \mathcal{B}%
_{0}\right) $ such that $T=G_{\nu _{T}}+L_{T}\times S^{2}$ over $\mathcal{D}%
^{3}(\mathcal{B}_{0}\times \mathbb{\hat{R}}^{3})$. If $T\in cart^{2,1}\left( 
\mathcal{B}_{0}\times S^{2}\right) $, then $\nu _{T}$ and $L_{T}$ are
uniquely defined by $T$ (see \cite{GMS6}, \cite{GMS7}, \cite{GMu}).

The extension of the Dirichlet energy associated with $S^{2}-$valued
morphological descriptor maps to $cart^{2,1}\left( \mathcal{B}_{0}\times
S^{2}\right) $ can be obtained by defining the energy on the space $%
cart^{2,1}\left( \mathcal{B}_{0}\times S^{2}\right) $ as 
\begin{equation*}
\mathcal{D}\left( T\right) :=\int_{\mathcal{B}_{0}}F(n,\overset{\rightarrow }%
{T})\text{ }d\left\Vert T\right\Vert ,
\end{equation*}%
where $F\left( n,\xi \right) $ is the polyconvex extension of the integrand 
\begin{equation*}
f\left( n,N\right) :=\left\{ 
\begin{array}{c}
\frac{1}{2}\left\vert N\right\vert ^{2}\text{ \ \ \ \ if }N^{\ast }n=0 \\ 
+\infty \text{ \ \ \ \ \ \ otherwise}%
\end{array}%
\right.
\end{equation*}%
to the space of $3-$vectors in $\Lambda _{3}\left( \mathbb{R}^{3}\times
S^{2}\right) $, while $\overset{\rightarrow }{T}:=\frac{dT}{d\left\Vert
T\right\Vert }$ is the Radon-Nykodim derivative of $T$ with respect to its
total variation. Precisely, $F\left( n,\xi \right) $\ is defined by 
\begin{eqnarray*}
F\left( n,\xi \right) &:&=\sup \left\{ \phi \left( \xi \right) \text{ }|%
\text{ }\phi :\Lambda _{3}\left( \mathbb{R}^{3}\times S^{2}\right)
\rightarrow \mathbb{R},\right. \\
&&\left. \phi \text{ linear},\text{ }\phi \left( M\left( N\right) \right)
\leq f\left( n,N\right) ,\forall \left( n,N\right) \right\}
\end{eqnarray*}%
One shows (see \cite{GMS7}, \cite{GMu}) that

\begin{description}
\item[(\emph{i})] 
\begin{equation}
\mathcal{D}\left( T\right) :=\frac{1}{2}\int_{\mathcal{B}_{0}}\left\vert
D\nu _{T}\right\vert ^{2}\text{ }dx+4\pi \mathbf{M}\left( L_{T}\right)
\label{101}
\end{equation}
if $T=G_{\nu _{T}}+L_{T}\times S^{2}\in cart^{2,1}\left( \mathcal{B}
_{0}\times S^{2}\right) $,

\item[(\emph{ii})] $\mathcal{D}\left( T\right) $ is lower semicontinuous due
to the convergence of currents with equibounded Dirichlet energies and

\item[(\emph{iii})] $\mathcal{D}\left( T\right) $ is the relaxed counterpart
of the Dirichlet integral for the convergence above.

\item[(\emph{iii})] (Closure) The class $cart^{2,1}\left( \mathcal{B}
_{0}\times S^{2}\right)$ is closed with respect to the convergence of
currents with equibounded masses and norms $\left\Vert \nu _{Tj}\right\Vert
_{W^{1,2}}$.
\end{description}

Existence of minimizers for (\ref{101}) in $cart^{2,1}\left( \mathcal{B}%
_{0}\times S^{2}\right) $ under Dirichlet boundary conditions and absence of
gap then follows. Such minimizers describe ground states of bodies with spin
structure in which the gross deformation is neglected.

The results described above extend to an existence theorem of minimizers for
an elastic body with spin structure admitting a decomposed energy of the type

\begin{equation}
\mathcal{E}\left( u,\nu _{T}\right) =\int_{\mathcal{B}_{0}}e_{E}\left(
x,u,\nu _{T},Du\right) \text{ }dx+\frac{1}{2}\int_{\mathcal{B}%
_{0}}\left\vert D\nu _{T}\right\vert ^{2}\text{ }dx+4\pi \mathbf{M}\left(
L_{T}\right) .  \label{EnSpin}
\end{equation}

\begin{theorem}
If $e_{E}$ satisfies (H1) and either (H2) or (H3), then the functional (\ref%
{EnSpin}) admits minimizers in $W^{r,s}(\mathcal{B}_{0},\mathbb{\hat{R}}
^{3})\times cart^{2,1}(\mathcal{B}_{0}\times S^{2})$ under Dirichlet
conditions.
\end{theorem}

The theorem above is the main result of the present section, the details of
the proof are not specified here because they are implied directly. The
presence of the term $4\pi \mathbb{M}(L_T)$ in \eqref{EnSpin} is a
constitutive choice. It takes into account the energetic contribution of the
overall behavior of the regions of the body in which the disorder is so high
that the identification of a local orientation becomes meaningless, regions
that are here described by the one dimensional current $L_T$.

\subsection{The general case}

Here, previous remarks are generalized. The aim is to obtain existence
results allowing (\emph{i}) the interactions between the substructural
changes and gross deformation even at level of first gradients (taking also
into account the minors involving elements of \emph{both} $F$ and $N$) and,
contemporarily, (\emph{ii}) the possible localization of substructural
activity along lines, with a possible generation of local substructural
disorder.

Note that the energy density $e$ does not depend (at least as far as one may
imagine in common cases) on the product $u\left( x\right) D\nu \left(
x\right) $. In fact, the derivative of $e$ with respect to $u$ is the
representative of standard external body forces and it is not natural to
presume that the standard body forces depend on the relative changes in
material substructure from place to place.

Below it is convenient to consider the pair deformation-morphological
descriptor as a unique map $\left( u,\nu \right) :\mathcal{B}_{0}\rightarrow 
\mathbb{\hat{R}}^{3}\times \mathcal{M}$.

All minors of the matrix 
\begin{equation*}
\left( 
\begin{array}{c}
F \\ 
N%
\end{array}
\right)
\end{equation*}
are collected in $M(%
\begin{pmatrix}
F \\ 
N%
\end{pmatrix}%
)$. Define the set 
\begin{equation*}
\mathcal{I}:=\left\{ \left( \alpha ,\beta \right) |\alpha \in I\left(
k,3\right) ,\text{ }\beta \in I\left( k,3+N\right) ,\text{ }0\leq k\leq 3 
\text{ s. t.}\right.
\end{equation*}
\begin{equation*}
\left. \left\vert M_{\beta }^{\alpha }\left( 
\begin{array}{c}
F \\ 
N%
\end{array}
\right) \right\vert \leq e\left( x,u,F,\nu ,N\right) ,\text{ }\forall \left(
x,u,F,\nu ,N\right) \right\}
\end{equation*}
and let $\mathcal{J}\subset \mathcal{I}$ be defined by 
\begin{equation*}
\mathcal{J}:=\left\{ \left( \alpha ,\beta \right) |\alpha \in I\left(
k,3\right) ,\text{ }\beta \in I\left( k,3+N\right) ,\text{ }0\leq k\leq 3 
\text{ s.t. }\exists \text{ }r>1\text{ s.t.}\right.
\end{equation*}
\begin{equation*}
\left. \left\vert M_{\beta }^{\alpha }\left( 
\begin{array}{c}
F \\ 
N%
\end{array}
\right) \right\vert ^{r}\leq e\left( x,u,F,\nu ,N\right) ,\text{ }\forall
\left( x,u,F,\nu ,N\right) \right\} .
\end{equation*}

As an additional constitutive assumption on the energy, it is assumed here
that the energy density $e$ is such that 
\begin{equation*}
\left( \alpha ,\beta \right) \in \mathcal{I}\Longrightarrow \left( \alpha
^{\prime },\beta ^{\prime }\right) \in \mathcal{J},\text{ \ }\forall \left(
\alpha ^{\prime },\beta ^{\prime }\right) ,\text{ }\alpha ^{\prime }<\alpha ,%
\text{ \ \ }\beta ^{\prime }<\beta .
\end{equation*}

If $\left( u_{j},\nu _{j}\right) $ is a sequence of pairs
deformation-morphological descriptor which are equibounded in energy, then
for any $\alpha $\ and $\beta $\ the sequence 
\begin{equation*}
\left\{ M_{\beta }^{\alpha }\left( 
\begin{array}{c}
Du_{j} \\ 
D\nu _{j}%
\end{array}
\right) \right\}
\end{equation*}
is equibounded in $L^{1}\left( \mathcal{B}_{0}\right) $ if $\left( \alpha
,\beta \right) \in \mathcal{I}$\ and equibounded in some $L^{r}\left( 
\mathcal{B}_{0}\right) $, $r>1$, if $\left( \alpha ^{\prime },\beta ^{\prime
}\right) \in \mathcal{J}$. Consequently, by taking subsequences, $\left(
u_{j},\nu _{j}\right) \rightarrow \left( u,\nu \right) $ in $L^{1}$ and
there exist measures $\mu _{\beta }^{\alpha }$ in $\mathcal{B}_{0}\times 
\mathbb{\hat{R}}^{3}\times \mathbb{R}^{N}$ such that for any $\phi \in
C^{0}( \mathcal{B}_{0}\times \mathbb{\hat{R}}^{3}\times \mathbb{R}^{N})$ one
gets 
\begin{equation*}
\int_{\mathcal{B}_{0}}\phi \left( x\mathbf{,}u_{j}\left( x\right) ,\nu
_{j}\left( x\right) \right) M_{\beta }^{\alpha }\left( 
\begin{array}{c}
Du_{j} \\ 
D\nu _{j}%
\end{array}
\right) \text{ }dx\rightarrow
\end{equation*}
\begin{equation*}
\mathbf{\rightarrow }\int_{\mathcal{B}_{0}\times \mathbb{\hat{R}}^{3}\times 
\mathbb{R}^{N}}\phi \left( x\mathbf{,}u\left( x\right) ,\nu \left( x\right)
\right) \text{ }d\mu _{\beta }^{\alpha }\left( x,y,\nu \right)
\end{equation*}
for any $\left( \alpha ,\beta \right) \in \mathcal{I}$. Moreover, if $\left(
\alpha ,\beta \right) \in \mathcal{J}$, one also gets 
\begin{equation*}
\int_{\mathcal{B}_{0}\times \mathbb{\hat{R}}^{3}\times \mathbb{R}^{N}}\phi
\left( x\mathbf{,}u\left( x\right) ,\nu \left( x\right) \right) \text{ }d\mu
_{\beta }^{\alpha }\left( x,y,\nu \right) =
\end{equation*}
\begin{equation*}
=\int_{\mathcal{B}_{0}}\phi \left( x\mathbf{,}u\left( x\right) ,\nu \left(
x\right) \right) M_{\beta }^{\alpha }\left( 
\begin{array}{c}
Du \\ 
D\nu%
\end{array}
\right) \text{ }dx\mathbf{.}
\end{equation*}

For each $j$, the previous measures can be collected in a vector-valued
measure, or, better, in the \emph{semi-currents} 
\begin{equation*}
G_{\left( u_{j},\nu _{j}\right) }:=\int_{\mathcal{B}_{0}}<\omega \left( x%
\mathbf{,}u_{j}\left( x\right) ,\nu _{j}\left( x\right) \right) ,M\left( 
\begin{array}{c}
Du_{j} \\ 
D\nu _{j}%
\end{array}%
\right) >\text{ }dx\mathbf{,}
\end{equation*}%
and 
\begin{equation*}
G_{\left( u,\nu \right) }:=\int_{\mathcal{B}_{0}}<\omega \left( x\mathbf{,}%
u\left( x\right) ,\nu \left( x\right) \right) ,M\left( 
\begin{array}{c}
Du \\ 
D\nu%
\end{array}%
\right) >\text{ }dx\mathbf{,}
\end{equation*}%
defined over the space 
\begin{equation*}
\mathcal{D}^{3,\mathcal{I}}:=\left\{ \omega =\sum_{\substack{ \beta :=\left(
\beta _{1},\beta _{2}\right)  \\ \left( \alpha ,\beta \right) \in \mathcal{I}
}}\omega _{\alpha ,\beta _{1},\beta _{2}}\left( x,u,\nu \right) \text{ }%
dx^{\alpha }\wedge du^{\beta _{1}}\wedge dz^{\beta _{2}}\right\} ,
\end{equation*}%
of $3-$forms on $\mathcal{B}_{0}\times \mathbb{\hat{R}}^{3}\times \mathbb{R}%
^{N}$. It follows that

\begin{description}
\item[(\emph{i})] the $G_{\left( u_{j},\nu _{j}\right) }$'s have equibounded
masses,

\item[(\emph{ii})] $G_{\left( u_{j},\nu _{j}\right) }\rightarrow
T:=G_{\left( u,\nu \right) }+S$,\ over $\mathcal{D}^{3,\mathcal{I}},$

\item[(\emph{iii})] the component $S^{\alpha \beta }\left( \phi \right)
:=S\left( \phi \left( x,u,\nu \right) \text{ }dx^{\alpha }\wedge du^{\beta
_{1}}\wedge dz^{\beta _{2}}\right) $ is identically zero if $\left( \alpha
,\beta \right) \in \mathcal{J}$.
\end{description}

As appropriate extension of the energy $e=e\left( x,y,\nu ,F,N\right) $, one
considers its polyconvex form%
\begin{equation*}
Pe\left( x,u,\nu ,\xi \right) =\sup \left\{ \phi \left( \xi \right) |\phi
\in Hom(\Lambda _{3}(\mathbb{R}^{3}\times \mathbb{\hat{R}}^{3}\times \mathbb{%
R}^{N}),\mathbb{R)},\text{ s.t.}\right.
\end{equation*}%
\begin{equation*}
\left. \phi \left( M\left( 
\begin{array}{c}
F \\ 
N%
\end{array}%
\right) \right) \leq e\left( x,u,F,\nu ,N\right) ,\text{ }\forall \left(
x,u,F,\nu ,N\right) \right\} .
\end{equation*}%
Note that, differently from the existence results in previous sections, here
the polyconvexification of the energy accounts for all minors of the matrix $%
\left( 
\begin{array}{c}
F \\ 
N%
\end{array}%
\right) $.

It is rather simple to show that $Pe$ is (\emph{i}) l.s.c. in $\xi $ for any
fixed $\left( x,u,\nu \right) $, (\emph{ii}) positively homogeneous of
degree 1 in $\xi $ and (\emph{iii}) $Pe\left( x,u,\nu ,\xi \right) \geq
\left\Vert \xi \right\Vert $. Then it is possible to extend the energy
functional to vector-valued measures $T=\left\{ T_{\beta }^{\alpha }\right\}
_{\left( \alpha ,\beta \right) \in \mathcal{I}}$ in the direct product $%
\mathbb{R}^{3}\times \mathbb{\hat{R}}^{3}\times \mathbb{R}^{N}$ by putting 
\begin{equation*}
\mathcal{F}\left( T\right) :=\int Pe(x,u,\nu ,\overset{\rightarrow }{T})%
\text{ }d\left\Vert T\right\Vert .
\end{equation*}%
A classical semicontinuity result of Reshetnyak (\cite{R1}, \cite{R2}, \cite%
{R3}) states that $\mathcal{F}$ is semicontinuous with respect to the weak
convergence of measures under the further assumption 
\begin{equation*}
Pe\left( x,u,\nu ,\xi \right) \text{ is l.s.c. in }\left( x,u,\nu ,\xi
\right) .
\end{equation*}%
Then, existence of minimizers in a class of semi-currents closed under the
weak convergence of measures follows trivially.

Such a general program has to be completed by analyzing the current $S$ in
the item (\emph{ii}) above, by computing explicitly the integral functional
and then discussing the possible absence of gap phenomenon. This program as
been developed in \cite{GMS7}, \cite{GMu} in the special case of the
Dirichlet energy $\frac{1}{2}\int_{\mathcal{B}_{0}}\left\vert D\nu
\right\vert ^{2}$ $dx$, with $\nu :\mathcal{B}_{0}\rightarrow \mathcal{M}$,
where $\mathcal{M}$ is a compact, oriented, Riemannian manifold $\mathcal{M}$
of dimension $\geq 2$. The general case is still open.

\section{Balance of standard and substructural actions}

The deduction of Euler-Lagrange equations for (\ref{Gen}) points out the
nature of the interactions involved in the mechanical behavior of complex
bodies and also the nature of their integral versions (see discussions in 
\cite{M02}, \cite{dFM}, \cite{M06}, \cite{M07}). Here the meaning of
Euler-Lagrange equations associated with irregular minimizers of the energy
of complex bodies and the conditions under which they exist are discussed.

\subsection{Euler-Lagrange equation: $C^{1}-$minimizers}

The condition 
\begin{equation}
\delta \mathcal{E}\left( u,\nu \right) =0,  \label{VarFun}
\end{equation}%
where $\delta $\ indicates first variation, characterizes the equilibrium.

For evaluating the first variation of $\mathcal{E}\left( u,\nu \right) $ it
is not necessary to embed the manifold of substructural shapes in some
linear space. $\mathcal{M}$\ is then considered abstract as in the original
format of the mechanics of complex bodies.

Assume first that $\mathcal{E}$ admits minimizers of class $C^{1}$. To
define variations over $\mathcal{M}$, it is useful to make use of fields of
the type $\upsilon :\mathcal{B}_{0}\rightarrow T\mathcal{M}$, with $\upsilon
\left( x\right) \in T_{\nu \left( x\right) }\mathcal{M}$, belonging to the
class $C_{c}^{1}\left( \mathcal{B}_{0},T\mathcal{M}\right) $ . For any $x\in 
\mathcal{B}_{0}$, here $\upsilon $ is taken such that $\upsilon =\frac{d}{%
d\varepsilon }\nu _{\varepsilon }\left\vert _{\varepsilon =0}\right.
:=\upsilon \left( x\right) $ in any local chart, being $\nu _{\varepsilon }$
a generic smooth curve $\left( -1,1\right) \ni \varepsilon \mapsto \nu
_{\varepsilon }\in \mathcal{M}$ crossing $\nu $ when $\varepsilon =0$.

Define 
\begin{equation*}
C_{\bar{u}}^{1}(\mathcal{B}_{0},\mathbb{\hat{R}}^{3}):=\left\{ u\in C_{\bar{%
u }}^{1}(\mathcal{B}_{0},\mathbb{\hat{R}}^{3})|u=\bar{u}\text{ on }\partial 
\mathcal{B}_{0}\right\} ,
\end{equation*}
\begin{equation*}
C_{\bar{\nu}}^{1}\left( \mathcal{B}_{0},\mathcal{M}\right) :=\left\{ \nu \in
C_{\bar{\nu}}^{1}\left( \mathcal{B}_{0},\mathcal{M}\right) |\nu =\bar{\nu} 
\text{ on }\partial \mathcal{B}_{0}\right\} ,
\end{equation*}
where $\bar{u}$\ and $\bar{\nu}$ are the boundary data in the Dirichlet
problem considered here.

If the energy $\mathcal{E}\left( u,\nu \right) $ attains a minimum at the
pair $\left( u,\nu \right) \in C_{\bar{u}}^{1}(\mathcal{B}_{0},\mathbb{\hat{%
R }}^{3})\times C_{\bar{\nu}}^{1}\left( \mathcal{B}_{0},\mathcal{M}\right) $%
, for $\varepsilon \in \left( -1,1\right) $ and each $h\in C_{c}^{1}(%
\mathcal{B }_{0},\mathbb{\hat{R}}^{3})$, the function $\varepsilon \mapsto 
\mathcal{E} \left( u+\varepsilon h,\nu _{\varepsilon }\right) $ attains a
minimum at $\varepsilon =0$.

The \emph{first variation} $\delta _{h,\upsilon }\mathcal{E}$ of $\mathcal{E}
$ \emph{from the ground state along the direction} $\left( h,\upsilon
\right) $ is then defined naturally by 
\begin{equation*}
\delta _{h,\upsilon }\mathcal{E}\left( u,\nu \right) :=\frac{d}{d\varepsilon 
}\mathcal{E}\left( u+\varepsilon h,\nu _{\varepsilon }\right) \left\vert
_{\varepsilon =0}\right. .
\end{equation*}
At $\left( u,\nu \right) $ the first variation \ of the energy then vanishes
along any direction $\left( h,\upsilon \right) $. Euler-Lagrange equations
then follows from the calculation of the first variation:

\begin{theorem}
Let the pair $\left( u,\nu \right) $\ be a minimizer for $\mathcal{E}$.
Then, (i) for any $h\in C_{c}^{1}\left( \mathcal{B}_{0},\mathbb{R}
^{3}\right) \ $and for any $\upsilon \in C_{c}^{1}\left( \mathcal{B}_{0},T 
\mathcal{M}\right) $, with $\upsilon \left( x\right) \in T_{\nu \left(
x\right) }\mathcal{M}$,\ the map $\varepsilon \mapsto \mathcal{E}\left(
u+\varepsilon h,\nu _{\varepsilon }\right) $ is differentiable and the pair $%
\left( u,\nu \right) $ satisfies the weak form of Euler-Lagrange equations 
\begin{equation}
\int_{\mathcal{B}_{0}}\left( -b\cdot h+P\cdot Dh+\zeta \cdot \upsilon + 
\mathcal{S}\cdot D\upsilon \right) \text{ }dx=0,  \label{WeakEL}
\end{equation}
(ii) if $\left( u,\nu \right) \in C^{2}\left( \mathcal{B}_{0},\mathbb{R}
^{3}\right) \times C^{2}\left( \mathcal{B}_{0},\mathcal{M}\right) $, then ( %
\ref{WeakEL}) is equivalent to the strong form 
\begin{equation}
DivP+b=0,  \label{Cau}
\end{equation}
\begin{equation}
Div\mathcal{S}-\zeta =0\text{ \ \ in }T_{\nu }^{\ast }\mathcal{M}.
\label{Cap}
\end{equation}
\end{theorem}

In the equations above, $P\left( x\right) :=\partial _{F}e\in
Hom(T_{x}^{\ast }\mathcal{B}_{0},T_{u\left( x\right) }^{\ast }\mathcal{B)}
\simeq \mathbb{R}^{3}\otimes \mathbb{\hat{R}}^{3}$ is the first
Piola-Kirchhoff stress, $b\left( x\right) :=-\partial _{u}e\in T_{x}^{\ast } 
\mathcal{B}\simeq \mathbb{R}^{3}$ the vector of standard body forces, $%
\mathcal{S}\left( x\right) :=\partial _{D\nu }e\in Hom(T_{x}^{\ast }\mathcal{%
\ B}_{0},T_{\nu \left( x\right) }^{\ast }\mathcal{M)}\simeq \mathbb{R}
^{3}\otimes T_{\nu \left( x\right) }^{\ast }\mathcal{M}$ the microstress
measuring constant interactions between neighboring material elements due to
substructural changes, $\zeta \left( x\right) :=\partial _{\nu }e\in T_{\nu
\left( x\right) }^{\ast }\mathcal{M}$. In particular, by considering $e$
decomposed additively in internal $e^{i}\left( x,Du,\nu ,D\nu \right) $ and
external $e^{e}\left( u,\nu \right) $ components, $\zeta $\ splits in the
sum $\zeta =z-\beta $, where $z\left( x\right) :=\partial _{\nu }e^{i}\in
T_{\nu \left( x\right) }^{\ast }\mathcal{M}$ is the self-action within the
generic material element due to substructural changes inside it while $\beta
\left( x\right) :=-\partial _{\nu }e^{e}\in T_{\nu \left( x\right) }^{\ast } 
\mathcal{M}$ represents external direct body actions on the substructure (a
paradigmatic example is the one of electric fields acting on the
polarization structure in ferroelectrics). Take note that the term $\partial
_{D\nu }e\cdot D\upsilon $ can be considered as the derivative $\frac{d}{
d\varepsilon }e\left( x,u+\varepsilon h,F+\varepsilon Dh,\nu _{\varepsilon
},N+\varepsilon D\upsilon \right) \left\vert _{\varepsilon =0}\right. $
since $N:=D\nu \left( x\right) $ belongs to a linear space. The same meaning
cannot be attributed to the term $\partial _{\nu }e\cdot \upsilon $\ because 
$\mathcal{M}$\ is not a linear space. The dot denotes the natural pairing
between dual spaces.

\subsection{Direct representation of standard and substructural actions:
invariance and balance}

Really the balance of actions involved in the mechanics of complex bodies
has the same structure of (\ref{Cau}), (\ref{Cap}), independently of
constitutive issues introduced in specifying the functional dependence of
the energy on the state variables in the variational setting considered
here. Equation (\ref{Cau}) is Cauchy balance of standard forces while (\ref%
{Cap}) is Capriz balance of substructural actions. A special case of (\ref%
{Cap}) is Ginzburg-Landau equation.

To obtain balance equations from the sole direct representation of standard
and substructural actions two tools are necessary: (\emph{i}) a class $%
\mathfrak{P}$ of subsets $\mathfrak{b}$ of $\mathcal{B}_{0}$\ with
non-vanishing volume and the same geometrical regularity of $\mathcal{B}_{0}$
\ itself, subsets\ called \emph{parts}, and (\emph{ii}) vector fields $h\in
C(\mathcal{B}_{0},\mathbb{\hat{R}}^{3})$ and $\upsilon \in C\left( \mathcal{B%
}_{0},T\mathcal{M}\right) $. Only the power of actions is defined here,
without paying attention to constitutive issues.

Given the pair $\tau :=\left( u,\nu \right) $, any \emph{power} along $%
\left( u,\nu \right) $\ is such a map $\mathcal{P}:\mathfrak{P}\left( 
\mathcal{B}_{0}\right) \times T\mathfrak{G}\rightarrow \mathbb{R}^{+}$ that $%
\mathcal{P}\left( \cdot ,\tau ,\dot{\tau}\right) $ is additive over disjoint
parts and $\mathcal{P}\left( \mathfrak{b},\tau ,\cdot \right) $ is linear.

The basic point is the explicit representation of $\mathcal{P}$, that is the
representation of actions over the generic $\mathfrak{b}$. The usual
assumption is that the actions be of volume and contact nature, the latter
represented by means of appropriate stresses that are in this case the first
Piola-Kirchhoff stress $P$\ and the microstress $\mathcal{S}$, no matter
about their possible constitutive structure. External bulk actions are
represented by the standard covector $b$ of body forces and, at each $x$, by
an element of the cotangent space $T_{\nu \left( x\right) }^{\ast }\mathcal{M%
}$ , indicated by $\beta $. In this way, the power $\mathcal{P}_{\mathfrak{b}%
}^{ext}\left( h\mathbf{,}\upsilon \right) $\ exchanged by the generic $%
\mathfrak{b}$ with the rest of the body and the external environment, a
power measured over $\left( h\mathbf{,}\upsilon \right) $\ along $\left(
u,\nu \right) $, is represented by 
\begin{equation}
\mathcal{P}_{\mathfrak{b}}^{ext}\left( h\mathbf{,}\upsilon \right) :=\int_{%
\mathfrak{b}}\left( b\cdot h+\beta \cdot \upsilon \right) \text{ }%
dx+\int_{\partial \mathfrak{b}}\left( Pn\cdot h+\mathcal{S}n\cdot \upsilon
\right) \text{ }d\mathcal{H}^{2}\mathfrak{,}
\end{equation}%
where $d\mathcal{H}^{2}$\ is the two-dimensional Hausdorff measure on $%
\partial \mathfrak{b}$, $n$ the normal to $\partial \mathfrak{b}$ in all
places in which it is defined, that is everywhere except a closed subset of $%
\partial \mathfrak{b}$\ with vanishing $\mathcal{H}^{2}$ measure. Here $b$, $%
\beta $, $P$ and $\mathcal{S}$ are not defined a priori as the derivatives
of the energy as energy does not come into play. Once $u$ and $\nu $\ are
given, one says only that, at each $x$, fields taking values $b\left(
x\right) \in T_{u\left( x\right) }^{\ast }\mathcal{B}\simeq \mathbb{\hat{R}}%
^{3}$, $\beta \left( x\right) \in T_{\nu \left( x\right) }^{\ast }\mathcal{M}
$ (bulk actions) and $P\left( x\right) \in Hom(T_{x}^{\ast }\mathcal{B}%
_{0},T_{u\left( x\right) }^{\ast }\mathcal{B)}\simeq \mathbb{R}^{3}\otimes 
\mathbb{\hat{R}}^{3}$, $\mathcal{S}\left( x\right) \in Hom(T_{x}^{\ast }%
\mathcal{B}_{0},T_{\nu \left( x\right) }^{\ast }\mathcal{M)}\simeq \mathbb{R}%
^{3}\otimes T_{\nu \left( x\right) }^{\ast }\mathcal{M}$ (contact
interactions) are defined.

\emph{Observers }are\emph{\ representations of the geometrical environments
necessary to describe the morphology of a body and its subsequent changes of
morphology} (see \cite{M07} for a series of questions related with this
definition).

Attention is focused here on semi-classical changes in observers, the ones
leaving invariant $\mathcal{B}_{0}$ and changing isometrically both $\mathbb{%
\hat{R}}^{3}$ and $\mathcal{M}$. The attribute `semi' refers to the
circumstance that $\mathcal{M}$ is taken into account in addition to the
ambient space. By considering the infinitesimal generators of the action of $%
\mathbb{\hat{R}}^{3}\ltimes SO\left( 3\right) $ over $\mathbb{\hat{R}}^{3}$
and of the same copy of $SO\left( 3\right) $ over $\mathcal{M}$, one defines 
$h^{\ast }:=h+c+q\times (u-u_{0})$, with $c\in \mathbb{\hat{R}}^{3}$ and $%
q\times \in so\left( 3\right) $ $u_{0}$ an arbitrary point in space, and $%
\upsilon ^{\ast }:=\upsilon +\mathcal{A}q$, where $\mathcal{A}\left( \nu
\right) \in Hom(\mathbb{\hat{R}}^{3},T_{\nu }\mathcal{M)}$ so that $\mathcal{%
A}^{\ast }\left( \nu \right) \in Hom(T_{\nu }^{\ast }\mathcal{M},\mathbb{%
\hat{R}}^{3})$. Here $h$ and $\upsilon $ play the role of virtual rates.

\textbf{Axiom}. \emph{At equilibrium the power of external actions is
invariant under semi-classical changes in observers, that is }$\mathcal{P}_{ 
\mathfrak{b}}^{ext}\left( h,\upsilon \right) =\mathcal{P}_{\mathfrak{b}
}^{ext}\left( h^{\ast },\upsilon ^{\ast }\right) $ \emph{for any choice of} $%
\mathfrak{b}$, $c$ \emph{and} $q$.

The following theorem is immediate (see \cite{M02}, \cite{M06} for further
remarks):

\begin{theorem}
(i) If for any $\mathfrak{b}$\ the vector fields $x\mapsto Pn$ and $x\mapsto 
\mathcal{A}^{\ast }\mathcal{S}n$ are defined over $\partial \mathfrak{b}$
and are integrable there, the integral balances of actions on $\mathfrak{b}$
\ hold: 
\begin{equation*}
\int_{\mathfrak{b}}b\text{ }dx+\int_{\partial \mathfrak{b}}Pn\text{ }d 
\mathcal{H}^{2}=0,
\end{equation*}
\begin{equation*}
\int_{\mathfrak{b}}\left( \left( u-u_{0}\right) \times b+\mathcal{A}^{\ast
}\beta \right) \text{ }dx+\int_{\partial \mathfrak{b}}\left( \left(
u-u_{0}\right) \times Pn+\mathcal{A}^{\ast }\mathcal{S}n\right) \text{ }d 
\mathcal{H}^{2}=0.
\end{equation*}
(ii) Moreover, if the tensor fields $x\mapsto P$ and $x\mapsto \mathcal{S}$
are of class $C^{1}\left( \mathcal{B}_{0}\right) \cap C^{0}\left( \mathcal{\ 
\bar{B}}_{0}\right) $ then 
\begin{equation*}
DivP+b=0
\end{equation*}
and there exist a covector field $x\mapsto z\in T_{\nu \left( x\right) } 
\mathcal{M}$ such that 
\begin{equation*}
skw\left( PF^{\ast }\right) =\mathsf{e}\left( \mathcal{A}^{\ast }z+\left( D 
\mathcal{A}^{\ast }\right) \mathcal{S}\right)
\end{equation*}
and 
\begin{equation*}
Div\mathcal{S}-z+\beta =0,
\end{equation*}
with $z=z_{1}+z_{2}$, $z_{2}\in Ker\mathcal{A}^{\ast }$.
\end{theorem}

Above \textsf{e} is Ricci's tensor.

The integral balances in the theorem above are associated with the Killing
fields of the standard metric in the ambient space $\mathbb{\hat{R}}^{3}$.
In general, a pure integral balance of substructural actions does not make
sense because it would involve integrands taking values on $T^*\mathcal{M}$
which is not a linear space.

Substructural interactions appear in the integral balance of moments which
is not standard. Their appearances do not imply that they are couples
(specifically micro-couples) due to the presence of the operator $\mathcal{A}%
^{\ast }$. In fact, only the products $\mathcal{A}^{\ast }\beta $ and $%
\mathcal{A}^{\ast }\mathcal{S}n$ are properly couples while $\mathcal{S}$
and $\mathcal{S}n$ do not.

\subsection{Irregular minimizers: horizontal variations}

Consider local minimizers in $\mathcal{W} _{r,s}^{d}$ of $\mathcal{E}\left(
u,\nu \right) $ (see Theorem 2). For the sake of simplicity, the lower bound 
\begin{equation*}
Pe\left( x,u,\nu ,M\left( F\right) \mathbf{,}N\right) \geq c_{1}\left(
\left\vert M\left( F\right) \right\vert ^{r}+\frac{\left\vert M\left(
F\right) \right\vert ^{\bar{r}}}{\left( \mathop{\rm det}\nolimits F\right) ^{%
\bar{r}-1}} +\left\vert N\right\vert ^{s}\right)
\end{equation*}
for some $r,\bar{r},s>1$ and $c_{1}>0$, is assumed to be satisfied by the
energy density. Consequently, the energy functional 
\begin{equation*}
\mathcal{E}\left( u,\nu \right) :=\int_{\mathcal{B}_{0}}Pe\left( x,u,\nu
,M\left( F\right) \mathbf{,}N\right) \text{ }dx,
\end{equation*}
is coercive over $dif^{r,\bar{r}}(\mathcal{B}_{0},\mathbb{\hat{R}}%
^{3})\times W^{1,s}\left( \mathcal{B}_{0},\mathcal{M}\right) $, where 
\begin{equation*}
dif^{r,\bar{r}}(\mathcal{B}_{0},\mathbb{\hat{R}}^{3}):=\left\{ u\in
dif^{r,1}(\mathcal{B}_{0},\mathbb{\hat{R}}^{3})\text{ }|\text{ }M\left( D 
\hat{u}\right) \in L^{\bar{r}}\left( \tilde{u}\left( \mathcal{B}_{0}\right)
\right) \right\} .
\end{equation*}
Take note that $\frac{\left\vert M\left( F\right) \right\vert ^{\bar{r}}}{
\left( \mathop{\rm det}\nolimits F\right) ^{\bar{r}-1}}$ is equal to $%
\left\vert M\left( D\hat{u} \right) \right\vert ^{r}$ where $\hat{u}:=u^{-1}$
in the sense of Lusin representatives.

Basically, the lower bound above means that the energy of the complex body
under examination is greater than the one of a fictious body in which
self-actions are absent so that substructural actions are only of contact
type (microstress).

Let $\phi \in C_{0}^{1}\left( \mathcal{B}_{0},\mathbb{R}^{3}\right) $ and
consider for for $\varepsilon $ sufficiently small, the diffeomorphism $\Phi
_{\varepsilon }\left( x\right) :=x+\varepsilon \phi \left( x\right) $ from $%
\mathcal{B}_{0}$ into itself, diffeomorphisms that leave unchanged $\partial 
\mathcal{B}_{0}$. Consequently, for $\left\vert \varepsilon \right\vert
<\varepsilon _{0}$, with $\varepsilon _{0}$ fixed, one gets 
\begin{equation*}
u_{\varepsilon }\left( x\right) :=u\left( \Phi _{\varepsilon }^{-1}\left(
x\right) \right) \in dif^{r,\bar{r}}(\mathcal{B}_{0},\mathbb{\hat{R}}^{3}),
\end{equation*}%
\begin{equation*}
\nu _{\varepsilon }\left( x\right) :=\nu \left( \Phi _{\varepsilon
}^{-1}\left( x\right) \right) \in W^{1,s}\left( \mathcal{B}_{0},\mathcal{M}%
\right) .
\end{equation*}

The map $\Phi _{\varepsilon }$ implies also the transformations 
\begin{equation*}
F\rightarrow F_{\varepsilon }=F_{\varepsilon }\left( x\right) :=FD\Phi
_{\varepsilon }^{-1}\text{ \ and\ \ }N\rightarrow N_{\varepsilon
}=N_{\varepsilon }\left( x\right) :=ND\Phi _{\varepsilon }^{-1}.
\end{equation*}

In this way, one contructs a map 
\begin{equation}
\varepsilon \rightarrow \mathcal{E}\left( u_{\varepsilon },\nu _{\varepsilon
}\right) :=\int_{\mathcal{B}_{0}}Pe\left( x,u_{\varepsilon },\nu
_{\varepsilon },M\left( F_{\varepsilon }\right) \mathbf{,}N_{\varepsilon
}\right) \text{ }dx.  \label{map-eps}
\end{equation}%
For the sake of simplicity, assume that $Pe$ is differentiable and 
\begin{equation}
\left\vert Pe\right\vert \text{ \ , \ }\left\vert \partial _{x}Pe\right\vert
\leq c\left( \left\vert M\left( F\right) \right\vert ^{r}+\frac{\left\vert
M\left( F\right) \right\vert ^{\bar{r}}}{\left( \mathop{\rm det}\nolimits
F\right) ^{\bar{r}-1}}+\left\vert N\right\vert ^{s}\right) ,  \label{UB1}
\end{equation}%
\begin{equation}
\left\vert \partial _{M\left( F\right) }Pe\right\vert \leq c\left(
\left\vert M\left( F\right) \right\vert ^{r-1}+\frac{\left\vert M\left(
F\right) \right\vert ^{\bar{r}-1}}{\left( \mathop{\rm det}\nolimits F\right)
^{\bar{r}-1}}+\left\vert N\right\vert ^{\left( 1-\frac{1}{\bar{r}}\right)
s}\right) ,  \label{UB2}
\end{equation}%
\begin{equation}
\left\vert \partial _{N}Pe\right\vert \leq c\left( \left\vert M\left(
F\right) \right\vert ^{\left( 1-\frac{1}{s}\right) r}+\left( \frac{%
\left\vert M\left( F\right) \right\vert ^{\bar{r}}}{\left( \mathop{\rm det}%
\nolimits F\right) ^{\bar{r}-1}}\right) ^{\left( 1-\frac{1}{s}\right)
}+\left\vert N\right\vert ^{s-1}\right) .  \label{UB3}
\end{equation}%
Then, by Lebesgue's differentiation theorem, the map $\varepsilon
\rightarrow \mathcal{E}\left( u_{\varepsilon },\nu _{\varepsilon }\right) $
is differentiable at zero with derivatives bounded in $L^{1}$. As a
consequence, the theorem below folows.

\begin{theorem}
If $Pe\left( x,u,\nu ,M\left( F\right) \mathbf{,}N\right) $ satisfies (\ref%
{UB1})-(\ref{UB3}) above, for a local minimizer $\left( u,\nu \right) $ of $%
\mathcal{E}\left( u,\nu \right) $ in $\mathcal{W}_{r,s}^{d}$ one gets 
\begin{equation*}
F^{\ast }\partial _{F}e\left( x,u,\nu ,F\mathbf{,}N\right) \in L^{1}\left( 
\mathcal{B}_{0}\right) ,
\end{equation*}%
\begin{equation*}
N^{\ast }\partial _{N}e\left( x,u,\nu ,F\mathbf{,}N\right) \in L^{1}\left( 
\mathcal{B}_{0}\right) ,
\end{equation*}%
\begin{equation}
\int_{\mathcal{B}_{0}}\mathbb{P}\cdot D\phi \text{ }dx+\int_{\mathcal{B}%
_{0}}\partial _{x}e\cdot \phi \text{ }dx=0\text{ \ }\forall \phi \in
C_{0}^{1}\left( \mathcal{B}_{0},\mathbb{R}^{3}\right) ,  \label{WCon}
\end{equation}%
where $\mathbb{P}\left( x\right) \in Aut\left( \mathbb{R}^{3}\right) $ is
the extended Hamilton-Eshelby tensor defined by 
\begin{equation}
\mathbb{P}:=eI-F^{\ast }P-N^{\ast }\mathcal{S},  \label{HE}
\end{equation}%
that is 
\begin{equation*}
Div\mathbb{P-}\partial _{x}e=0,
\end{equation*}%
in distributional sense.
\end{theorem}

In fact, by using Binet formula and Young inequality (namely $ab\leq \frac{
a^{r}}{r}+\frac{b^{\bar{r}}}{\bar{r}}$, $\frac{1}{r}+\frac{1}{\bar{r}}=1$),\
it is easy to prove that 
\begin{equation*}
\sup_{\left\vert \varepsilon \right\vert <\varepsilon _{0}}\frac{d}{
d\varepsilon }\left( Pe\left( \phi \left( x\right) ,u_{\varepsilon },\nu
_{\varepsilon },M\left( F_{\varepsilon }\right) \mathbf{,}N_{\varepsilon
}\right) \mathop{\rm det}\nolimits D\Phi _{\varepsilon }\right) \in
L^{1}\left( \mathcal{B} _{0}\right) .
\end{equation*}
Precisely, Binet formula 
\begin{equation*}
M_{\alpha }^{\beta }\left( F_{\varepsilon }\right) =\sum_{\left\vert \gamma
\right\vert =\left\vert \beta \right\vert }M_{\gamma }^{\beta }\left(
F_{\varepsilon }\right) M_{\alpha }^{\gamma }\left( D\Phi _{\varepsilon
}^{-1}\right)
\end{equation*}
yields 
\begin{equation*}
\left\vert M\left( F_{\varepsilon }\right) \right\vert \text{\ },\text{\ }
\left\vert \frac{d}{d\varepsilon }M\left( F_{\varepsilon }\right)
\right\vert \leq c\left\vert M\left( F\right) \right\vert .
\end{equation*}
Moreover, one has 
\begin{equation*}
\left\vert N_{\varepsilon }\right\vert \text{\ },\text{\ }\left\vert \frac{d 
}{d\varepsilon }N_{\varepsilon }\right\vert \leq c\left\vert N\right\vert ,
\end{equation*}
so that 
\begin{equation*}
\left\vert \frac{d}{d\varepsilon }\left( Pe\left( \phi \left( x\right)
,u_{\varepsilon },\nu _{\varepsilon },M\left( F_{\varepsilon }\right) 
\mathbf{,}N_{\varepsilon }\right) \mathop{\rm det}\nolimits D\Phi
_{\varepsilon }\right) \right\vert \leq
\end{equation*}
\begin{equation*}
\leq c\left\{ \left\vert \partial _{x}Pe\right\vert +\left\vert \partial
_{M\left( F\right) }Pe\cdot \frac{d}{d\varepsilon }M\left( F_{\varepsilon
}\right) \right\vert +\left\vert \partial _{N}Pe\cdot \frac{d}{d\varepsilon }
N_{\varepsilon }\right\vert \right\} \leq
\end{equation*}
\begin{equation*}
\leq c(\left\vert M\left( F\right) \right\vert ^{r}+\frac{\left\vert M\left(
F\right) \right\vert ^{\bar{r}}}{\left( \mathop{\rm det}\nolimits F\right) ^{%
\bar{r}-1}} +\left\vert N\right\vert ^{s})+
\end{equation*}
\begin{equation*}
+c(\left\vert M\left( F\right) \right\vert ^{r-1}+\frac{\left\vert M\left(
F\right) \right\vert ^{\bar{r}-1}}{\left( \mathop{\rm det}\nolimits F\right)
^{\bar{r}-1}} +\left\vert N\right\vert ^{\left( 1-\frac{1}{\bar{r}}\right)
s})\left\vert M\left( F\right) \right\vert +
\end{equation*}
\begin{equation*}
+c(\left\vert M\left( F\right) \right\vert ^{\left( 1-\frac{1}{s}\right)
r}+\left( \frac{\left\vert M\left( F\right) \right\vert ^{\bar{r}}}{\left( %
\mathop{\rm det}\nolimits F\right) ^{\bar{r}-1}}\right) ^{\left( 1-\frac{1}{s%
}\right) }+\left\vert N\right\vert ^{s-1})\left\vert N\right\vert \leq
\end{equation*}
\begin{equation*}
\leq c(\left\vert M\left( F\right) \right\vert ^{r}+\frac{\left\vert M\left(
F\right) \right\vert ^{\bar{r}}}{\left( \mathop{\rm det}\nolimits F\right) ^{%
\bar{r}-1}} +\left\vert N\right\vert ^{s})
\end{equation*}
for all $\varepsilon $, $\left\vert \varepsilon \right\vert \leq \varepsilon
_{0}$.

Then, it follows that 
\begin{eqnarray}
0 &=&\frac{d}{d\varepsilon }\mathcal{E}\left( u_{\varepsilon },\nu
_{\varepsilon }\right) =\int_{\mathcal{B}_{0}}\frac{d}{d\varepsilon }\left(
Pe\left( \phi \left( x\right) ,u_{\varepsilon },\nu _{\varepsilon },M\left(
F_{\varepsilon }\right) \mathbf{,}N_{\varepsilon }\right) \mathop{\rm det}%
\nolimits D\Phi _{\varepsilon }\right) \left\vert _{\varepsilon =0}\right. 
\text{ }dx  \notag \\
&=&\int_{\mathcal{B}_{0}}\frac{d}{d\varepsilon }\left( e\left( \phi \left(
x\right) ,u_{\varepsilon },\nu _{\varepsilon },F_{\varepsilon }\mathbf{,}%
N_{\varepsilon }\right) \mathop{\rm det}\nolimits D\Phi _{\varepsilon
}\right) \left\vert _{\varepsilon =0}\right. \text{ }dx  \label{der}
\end{eqnarray}%
The derivative at $\varepsilon =0$ under the integral sign remains to be
computed. Since 
\begin{equation*}
\Phi _{\varepsilon }\left( x\right) =x+\varepsilon \phi \left( x\right) ,
\end{equation*}%
\begin{equation*}
D\Phi _{\varepsilon }^{-1}\left( \Phi _{\varepsilon }\left( x\right) \right)
=I-\varepsilon D\phi \left( x\right) +o\left( \varepsilon ^{2}\right) \text{
\ \ as }\varepsilon \rightarrow 0,
\end{equation*}%
\begin{equation*}
\mathop{\rm det}\nolimits D\Phi _{\varepsilon }\left( x\right)
=1+\varepsilon Div\Phi _{\varepsilon }\left( x\right) +o\left( \varepsilon
^{2}\right) \text{ \ \ as }\varepsilon \rightarrow 0,
\end{equation*}%
uniformly with respect to $x$ ($I$ indicates the unit second rank tensor),
the derivative in (\ref{der})\ implies (\ref{WCon}).

Theorem 7 extends to complex bodies a companion result for simple elastic
bodies in \cite{GMS7}. The extended Hamilton-Eshelby tensor $\mathbb{P}$ has
been introduced in \cite{M02} (see also \cite{dFM}) with reference to smooth
minimizers. Here the configurational balance involving $\mathbb{P}$ is
extended to irregular minimizers.

Actually, as pointed out by \cite{B2} for non-linear elasticity of simple
bodies, the differentiability of the map $\varepsilon \longmapsto \mathcal{E}%
\left( u_{\varepsilon },\nu _{\varepsilon }\right) $ in (\ref{map-eps})
holds actually under the weaker energetic estimate%
\begin{equation*}
\left\vert \partial _{x}e\right\vert +\left\vert F^{\ast }\partial
_{F}e\right\vert +\left\vert N^{\ast }\partial _{N}e\right\vert \leq
c_{1}e+c_{2},
\end{equation*}%
where $e$ and its derivatives are calculated in $\left( x,u,\nu ,F\right) $.
Such an estimate has been used in \cite{FM} for a delicate analysis of
evolution problems in rate-independent models of non-conservative processes
in classes of bodies.

In the case of bodies with spin structure described by $\nu :\mathcal{B}%
_{0}\rightarrow S^{2}$ and admitting a partially decomposed energy with
concentration on a line, precisely an energy of the form 
\begin{equation*}
\mathcal{E}\left( u,\nu _{T}\right) =\int_{\mathcal{B}_{0}}e_{E}\left(
x,u,\nu _{T},F\right) \text{ }dx+\frac{1}{2}\int_{\mathcal{B}_{0}}\left\vert
D\nu _{T}\right\vert ^{2}\text{ }dx+4\pi \mathbf{M}\left( L_{T}\right) ,
\end{equation*}%
with $L_{T}$\ a one-dimensional integer rectifiable current on $\mathcal{D}%
^{1}\left( \mathcal{B}_{0}\right) $, by using the technique adopted in the
proof of the theorem above, a special version of (\ref{WCon}) follows. It
accounts for the contribution of the concentration of energy on the line $%
L_{T}=\overrightarrow{T}\wedge \left\Vert L_{T}\right\Vert $, namely 
\begin{equation}
\int_{\mathcal{B}_{0}}\mathbb{P}\cdot D\phi \text{ }dx+\int_{\mathcal{B}%
_{0}}\partial _{x}e\cdot \phi \text{ }dx=4\pi \int \overset{\rightarrow }{T}%
\otimes \overset{\rightarrow }{T}\cdot D\phi \text{ }d\left\Vert
L_{T}\right\Vert  \label{WConW}
\end{equation}%
for any $\phi \in C_{0}^{1}\left( \mathcal{B}_{0},\mathbb{R}^{3}\right) $.
The integral balance (\ref{WConW}) is unusual. In the case of homogeneous
bodies, (\ref{WConW}) implies the the internal `power' of the extended
Hamilton-Eshelby stress is determined only by the interactions along the
line of concentration of energy. Equation (\ref{WConW}) extends also a
theorem in \cite{GMS2} to the case in which macroscopic deformations occur.

Consider maps $\bar{\phi}\in C_{c}^{1}(\mathbb{\hat{R}}^{3},\mathbb{\hat{R}}%
^{3})$ with $\bar{\phi}=0$ in a neighborhood of $u\left( \partial \mathcal{B}%
_{0}\right) $. For $\left\vert \varepsilon \right\vert <\varepsilon _{0}$,
with $\varepsilon _{0}$ fixed, the map $\Phi _{\varepsilon }\left( y\right)
=y+\varepsilon \bar{\phi}\left( y\right) $ is then a diffeomorphism from $%
\mathbb{\hat{R}}^{3}$ into $\mathbb{\hat{R}}^{3}$ , so the map $%
u_{\varepsilon }\left( x\right) :=u\left( x\right) +\varepsilon \bar{\phi}%
\left( u\left( x\right) \right) $ is a weak diffeomorphism in $dif^{r,\bar{r}%
}(\mathcal{B}_{0},\mathbb{\hat{R}}^{3})$. Since $\bar{\phi}=0$ in a
neighborhood of $\tilde{u}\left( \partial \mathcal{B}_{0}\right) $, all the $%
u_{\varepsilon }$ agree on $\partial \mathcal{B}_{0}$; moreover by chain
rule it follows that $F_{\varepsilon }=F_{\varepsilon }\left( x\right)
=F\left( x\right) +\varepsilon D_{u\left( x\right) }\bar{\phi}\left( u\left(
x\right) \right) F\left( x\right) $.

Define the Cauchy stress tensor as usual by 
\begin{equation*}
\mathbf{\sigma }\left( y\right) :=(\left( \mathop{\rm det}\nolimits F\right)
^{-1}\partial _{F}e\left( x,u,\nu ,F\mathbf{,}N\right) F^{\ast })\left(
y\right) \in Hom(T_{y}^{\ast }\mathcal{B},\mathbb{\hat{R}}^{3\ast })\simeq 
\mathbb{\hat{R}}^{3}\otimes \mathbb{\hat{R}}^{3}.
\end{equation*}%
Assume also that the energy density satisfies the inequality 
\begin{equation}
\left\vert \partial _{u}Pe\right\vert \leq c(\left\vert M\left( F\right)
\right\vert ^{r}+\frac{\left\vert M\left( F\right) \right\vert ^{\bar{r}}}{%
\left( \mathop{\rm det}\nolimits F\right) ^{\bar{r}-1}}+\left\vert
N\right\vert ^{s}).  \label{UB4}
\end{equation}%
Under this additional assumption and (\ref{UB1}), (\ref{UB2}), the map 
\begin{equation*}
\varepsilon \rightarrow \mathcal{E}\left( u_{\varepsilon },\nu \right)
:=\int_{\mathcal{B}_{0}}Pe\left( x,u_{\varepsilon },\nu ,M\left(
F_{\varepsilon }\right) \mathbf{,}N\right) \text{ }dx
\end{equation*}%
is differentiable at $\varepsilon =0$.

\begin{theorem}
Under conditions (\ref{UB1}), (\ref{UB2}) and (\ref{UB4}) above, for $\left(
u,\nu \right) $\ a minimizer in $W_{r,s}^{d}$ of $\mathcal{E}\left( u,\nu
\right) $, 
\begin{equation*}
\mathbf{\sigma }\in L_{loc}^{1}(\tilde{u}\left( \mathcal{B}_{0}\right) , 
\mathbb{\hat{R}}^{3}\otimes \mathbb{\hat{R}}^{3})
\end{equation*}
and 
\begin{equation*}
\int_{\tilde{u}\left( \mathcal{B}_{0}\right) }\mathbf{\sigma }\left(
y\right) \cdot D\bar{\phi}\left( y\right) \text{ }dy+\int_{\tilde{u}\left( 
\mathcal{B}_{0}\right) }b\left( y\right) \cdot \bar{\phi}\left( y\right) 
\text{ }dy=0,
\end{equation*}
for every\emph{\ }$\bar{\phi}\in C_{0}^{1}(\mathbb{\hat{R}}^{3},\mathbb{\hat{
R}}^{3})$ with\emph{\ }$\bar{\phi}=0$ in a neighborhood of $\tilde{u}\left(
\partial \mathcal{B}_{0}\right) $, with $\tilde{u}$ the Lusin representative
of $u$.
\end{theorem}

The proof follows by direct calculation.

Finally, consider smooth curves $\varepsilon \rightarrow \bar{\varphi}%
_{\varepsilon }\in Aut\left( \mathcal{M}\right) $, $\bar{\varphi}\in
C^{1}\left( \mathcal{M}\right) $, and define 
\begin{equation*}
\nu _{\varepsilon }:=\bar{\varphi}_{\varepsilon }\left( \nu \right) \text{ \
\ , \ \ }\nu \in \mathcal{M}\text{.}
\end{equation*}%
Call also $\xi $ the derivative 
\begin{equation*}
\xi :=\frac{d}{d\varepsilon }\nu _{\varepsilon }\left\vert _{\varepsilon
=0}\right. .
\end{equation*}

Assume that the energy density satisfies the inequality 
\begin{equation}
\left\vert \partial _{\nu }Pe\right\vert \leq c\left( \left\vert M\left(
F\right) \right\vert ^{r}+\frac{\left\vert M\left( F\right) \right\vert ^{ 
\bar{r}}}{\left( \mathop{\rm det}\nolimits F\right) ^{\bar{r}-1}}+\left\vert
N\right\vert ^{s}\right) .  \label{UB5}
\end{equation}

Under the assumptions (\ref{UB1}), (\ref{UB3}) and (\ref{UB5})\ the map 
\begin{equation*}
\varepsilon \rightarrow \mathcal{E}\left( u,\nu _{\varepsilon }\right)
:=\int_{\mathcal{B}_{0}}Pe\left( x,u,\nu _{\varepsilon },M\left( F\right) 
\mathbf{,}N_{\varepsilon }\right) \text{ }dx
\end{equation*}
is differentiable at $\varepsilon =0$.

\begin{theorem}
Under conditions (\ref{UB1}), (\ref{UB3}) and (\ref{UB5}) above, for $u$ and 
$\nu $ minimizers in $W_{r,s}^{d}$ of $\mathcal{E}\left( u,\nu \right) $, 
\begin{equation*}
\mathcal{S}\in L^{1}\left( \mathcal{B}_{0},\mathbb{R}^{3\ast }\otimes
T^{\ast }\mathcal{M}\right)
\end{equation*}
and 
\begin{equation*}
\int_{\mathcal{B}_{0}}\mathcal{S}\left( x\right) \cdot D\xi \left( x\right) 
\text{ }dx+\int_{\mathcal{B}_{0}}(\beta-z)\left( x\right) \cdot \xi \left(
x\right) \text{ }dx=0,
\end{equation*}
for every $\xi \in C^{0}\left( T\mathcal{M}\right)$, with $(\beta-z)(x)\in
T^{\ast}_{\nu(x)}\mathcal{M}$.
\end{theorem}

\section{Taxonomy of special cases}

Without having the presumption to be a theory of everything, the framework
discussed above unifies, in fact, a very large class of models of condensed
matter physics. A list of special examples is presented below. It is not
exhaustive, of course, but an idea of the potentialities of thinking of
complex bodies in terms of maps between manifolds is given.

\begin{itemize}
\item \textbf{Liquid crystals in nematic phase: }In liquid crystals stick
molecules are dispersed in a ground fluid. They may arrange themselves in
various manners that characterize different phases. In nematic phase, the
stick molecules are ordered along prevailing directions but they do not have
distinct head and tail so that $\mathcal{M}$ is identified with the unit
sphere in $\mathbb{R}^{3}$, with the projective plane $P^{2}$. The generic
material element can be interpreted here as a patch of matter including a
family of stick molecules. The morphological descriptor $\nu $ is then an
indicator of the `prevailing' direction of the molecules. This point of view
has been introduced in \cite{E1}, \cite{E2} (see also \cite{Les}, \cite{DGP}%
). Oseen-Frank potential, recalled above, is the energy appropriate when one
forgets the gross motion. A second-rank tensor $\zeta \left( n\otimes n-%
\frac{1}{3}I\right) $, with $n\in S^{2}$, and $\zeta \in \left[ -\frac{1}{2}%
,1\right] $, can be also used as to account for details of the distribution
of the stick molecules. The scalar $\zeta $ indicates the \emph{degree of
orientation} (as defined in \cite{E3}). In this case, then, $\mathcal{M}%
=S^{2}\times \left[ -\frac{1}{2},1\right] $. Optical biaxiality can emerge
so that the symmetry of the molecules becomes that of a rectangular box and
two other scalar morphological descriptors are necessary: the \emph{degree
of prolation} and the \emph{degree of triaxiality}\ (see \cite{CB}).
Alternatively, one may select $\mathcal{M}$ coincident with the quotient
between the special unitary group $SU\left( 2\right) $ and the group of
quaternions.

\item \textbf{Liquid crystals in smectic phase}: In the smectic-A phase a
layered structure appears and the stick molecules tend to be aligned
orthogonally to the layer interface unless tilt occurs. Natural ingredients
for describing the smectic-A phase are the unit vector $n$ representing at
each point the local orientational order and a scalar function $\ell $
parametrizing the layers through its level sets. When tilt is absent and
single layers are compressible but at the gross scale there is
incompressibility, the energy density can be written as 
\begin{equation*}
e\left( \ell ,grad\text{ }\ell \right) =\frac{1}{2}k_{1}\left( \left\vert
grad\text{ }\ell \right\vert -1\right) ^{2}+\frac{1}{2}k_{2}\left( div\text{ 
}n\right) ^{2},
\end{equation*}%
with $k_{1}$\ and $k_{2}$\ material constants and the operators $grad$ and $%
div$ imply derivatives with respect to $u$. The term $\left( \left\vert grad%
\text{ }\ell \right\vert -1\right) ^{2}$ accounts for the compression of
layers while $\left( div\text{ }n\right) ^{2}$ describes the nematic phase
and is the first addendum of (three constant) Frank's potential (see \cite%
{C94}).

\item \textbf{Liquid crystals in cholesteric phase}: In cholesteric phase,
liquid crystals loose in a sense reflection symmetry (the one under the
action of $O\left( 3\right) \backslash SO\left( 3\right) $) and maintain $%
SO\left( 3\right) $-symmetry. An appropriate form of the energy density can
be found in \cite{E3}.

\item \textbf{Cosserat materials}: In the Cosserat's scheme each material
element is considered as a (small) \emph{rigid body} which can rotate
independently of the neighboring fellows. The manifold of substructural
shapes can be then identified with the unit sphere $S^{2}$ or the special
orthogonal group $SO\left( 3\right) $. Local contact couples exchanged
between adjacent parts of the body are power conjugated with local rotations
and are described by a couple stress tensor. The scheme of Cosserat
materials is a special case of multifield theories often used for direct
models of structural elements like beams, plates or shells \cite{ET} or for
composites reinforced with diffused small rigid fibers.

\item \textbf{Superfluid liquid helium}: The analysis of ground states of $%
^{3}He$ falls within the setting discussed above. The energy density is of
Ginzburg-Landau type. For $^{3}He$ in the dipole locked phase, $\mathcal{M}$
\ coincides simply with $SO\left( 3\right) $ (thus with $S^{2}$) and
Cosserat's scheme applies.

\item \textbf{Ferroelectrics}: To describe the local polarization of
crystalline cells in ferroelectrics, a vector is commonly selected within a
ball $B_{p}$ in $\mathbb{R}^{3}$, the radius of which is the maximum
polarization available in the material \cite{S}. In presence of an external
electric field acting over the body, the relevant polarization energy has to
be added to a Ginzburg-Landau-type decomposed energy density for matter
fields.

\item \textbf{Bodies with polymeric chains}: Various types of materials are
made of polymeric linear chains scattered in a melt (see, e.g., \cite{Lik}).
A simple natural descriptor of each single chain is an end-to-end
stretchable vector $\mathbf{r}$. To preserve a natural symmetry under the
transformation $\mathbf{r\rightarrow -r}$, the dyad $\mathbf{r\otimes r}$ is
used as morphological descriptor. Then, the manifold of substructural shapes 
$\mathcal{M}$ coincides with the (linear) space of symmetric tensors with
positive determinant $Sym^{+}\left( \mathbb{R}^{3},\mathbb{R}^{3}\right) $.
In this sense the representation falls within the class of affine bodies
mentioned below. The energy may be selected in various manners; in
particular, if the linear chains are `dilute', the energy does not depend on 
$D\nu $ while, when their are `dense' up to interacting through van der
Walls forces and/or entanglements, the dependence on $D\nu $ appears.

\item \textbf{Polyelectrolyte polymers and polymer stars}: Polyelectrolyte
polymers are characterized by the possible polarization of chains. In this
case, by adopting the notation above, the manifold of substructural shapes
can be selected as $\mathcal{M}=Sym^{+}\left( \mathbb{R}^{3},\mathbb{R}%
^{3}\right) \times B_{p}$. Moreover the chains may link with each other up
to form a star. In this case we may imagine to have $\mathcal{M}%
=Sym^{+}\left( \mathbb{R}^{3},\mathbb{R}^{3}\right) \times B_{p_{m}}\times
\left( 0,c\right) $, with $c>0$. Numbers in $\left( 0,c\right) $ describe
the \emph{radius of gyration} of the star.

\item \textbf{Bodies with affine structure}: In the special case in which
the manifold of substructural shape coincides with the linear space of
second-rank tensors, the substructure is called affine \cite{Mn}, \cite{Slaw}%
. The scheme is suitable to cover various cases such as the one of bodies
with dense polymeric linear chains discussed above or fullerene-reinforced
composites. A basic interest to mention this special model is that when
there is an internal constraint of the type $\nu =f\left( F\right) $, so
that the substructure becomes \emph{latent} in the sense of Capriz \cite{C85}%
. In this case the energy density reduces to the one of a second-grade
Cauchy body, namely $e=e\left( x,F,\nabla F\right) $.

\item \textbf{Porous and multi-phase bodies}: When pores are finely
scattered throughout a body, we may imagine that the generic material
element is a patch of matter with spherical voids and we can select the
morphological descriptor as a scalar indicating the \textit{void volume
fraction}. In this case $\mathcal{M}$ reduces to the interval of the real
axis $\left[ 0,1\right] $ (see \cite{NuC}). The scheme of porous materials
is useful when multiple phases (for example $m$ phases) coexist within a
body and phase transitions occur \cite{Fr}. The description can be refined
to account for deeper details in the rearrangement of phases. The combined
use of scalar and second-order tensor valued morphological descriptor fields
allows one to account for the re-orientation of martensitic variants, as
proposed in \cite{BP}.

\item \textbf{Quasi-periodic alloys}: Quasi-periodic atomic arrangements
exist in nature and characterize some specific classes of metallic alloys.
For example, in the prominent case of quasicrystals, the formation and
annihilation of atomic rearrangements is necessary to assure
quasi-periodicity, so that internal degrees of freedom appear within the
crystalline cells and describe the local atomic rearrangements (a process
called \emph{phason activity}). In all cases, the appropriate morphological
descriptor of the degrees of freedom inside `atomic cells' is a vector so
that $\mathcal{M}$ coincides with $\mathbb{R}^{3}$ (see \cite{M06}, \cite%
{HWD}).
\end{itemize}

\ \ \ \ \ \ 

\section{Ground states of thermodynamically stable quasicrystals}

A special class of quasi-periodic alloys is the one of quasicrystals in
which atomic clusters display symmetries incompatible with periodic tilings
in space, symmetries such as the icosahedral one in three-dimensions and the
penthagonal one in the plane (see, e.g., \cite{HWD}). Atomic rearrangements
assure quasiperiodicity in space by creating and annihilating (randomly) the
so-called worms, that are clusters of atoms with symmetry different from the
prevailing one. The local degrees of freedom associated with these atomic
changes are described by a vector $\nu \in \mathbb{R}^{3}$ in the
three-dimensional case. $\nu $ belongs to $\mathbb{R}^{3}$\ for
two-dimensional quasicrystals. Experiments show that the elastic energy of
quasicrystals does not depend on $\nu $ while it depends only on its spatial
gradient $N$ besides the gradient of macroscopic deformation. Moreover,
quasicrystals are characterized by a self-action of dissipative nature, that
is by dissipation inside each material element, a dissipation strictly
associated with substructural events (see \cite{M06}). Although this type of
tendence to material metastability, it has been shown experimentally that
quasicrystals may admit in some cases ground states (see \cite{TGAT} and
references therein). Moreover, the occurrence of ground states have been
also analyzed in the generic material element by looking directly to the
lattice behavior (see \cite{DH}, \cite{Mi}). Additionally, it has been shown
also that frustration between neighboring ground states may occur in special
conditions \cite{DDO}.

Here, the framework presented above is applied to analyze the existence of
ground states in quasicrystalline bodies from a macroscopic point of view.
The energy to be accounted for is then 
\begin{equation*}
\int_{\mathcal{B}_{0}}e\left( Du,D\nu \right) \text{ }dx
\end{equation*}%
and one tries to find minimizers for it in some functional space. The
appropriate constitutive constitutive choice of the functional environment
seem to be the space%
\begin{equation*}
\mathcal{W}_{r,2}:=\left\{ \left( u,\nu \right) |u\in dif^{r,1}(\mathcal{B}%
_{0},\mathbb{\hat{R}}^{3}),\text{ }\nu \in W^{1,2}\left( \mathcal{B}_{0},%
\mathbb{R}^{3}\right) \right\} .
\end{equation*}%
In this case the growth condition for the energy becomes%
\begin{equation*}
e\left( F\mathbf{,}N\right) \geq C_{1}\left( \left\vert M\left( F\right)
\right\vert ^{r}+\left\vert N\right\vert ^{2}\right) +\vartheta \left( %
\mathop{\rm det}\nolimits F\right) .
\end{equation*}%
It means that the energy grows faster than the one of an ideal quasicrystal
behaving isotropically and having an unlocked phase. Precisely, one says
that the atomic rearrangements occurring in quasicrystals are in a unlocked
phase when the energy has a quadratic dependence on $N$. Such a phase is the
only one existing in two-dimensional setting (see \cite{JS}) so that the
growth condition above has physical meaning in one, two and
three-dimensional ambient space.

As a consequence, once $e\left( F\mathbf{,}N\right) $ is substituted by its
polyconvex extension, existence of minimizers in $\mathcal{W}_{r,2}$\
follows as a consequence of Theorem 2.

Analogous analyses can be developed for other prominent cases of complex
bodies.

\ \ \ \ \ \ \ \ \ \ \ \ \ \ \ 

\textbf{Acknowledgements.} We thank Mariano Giaquinta and Alexander Mielke
for helpful discussions. PMM acknowledges the support of the Italian
National Group of Mathematical Physics (GNFM-INDAM) and of MIUR under the
grant 2005085973-"\emph{Resistenza e degrado di interfacce in materiali e
strutture}"-COFIN 2005. GM acknowledges the support of MIUR under the grant "%
\emph{Calcolo delle variazioni}"-COFIN 2004. The hospitality of the "Centro
di Ricerca Matematica Ennio De Giorgi" of the Scuola Normale Superiore at
Pisa is gratefully acknowledged.

\ \

\end{document}